\documentclass{article}

\usepackage{epsf,epsfig}

 \usepackage{graphicx}
\usepackage{graphicx, epsfig, amssymb} 
\usepackage{bm}
\usepackage{amsfonts}
\usepackage{amsmath}

\usepackage{amssymb}
\usepackage{graphicx}
\usepackage{epsfig}
\usepackage{amsfonts}
\usepackage{a4}
\usepackage{amsmath}

\newcommand{\ee}{\end{equation}}
\newcommand{\eea}{\end{eqnarray}}
\newcommand{\be}{\begin{equation}}
\newcommand{\bea}{\begin{eqnarray}}

\def\dalemb#1#2{{\vbox{\hrule height .#2pt
        \hbox{\vrule width.#2pt height#1pt \kern#1pt
                \vrule width.#2pt}
        \hrule height.#2pt}}}

\def\0{{\sst{(0)}}}
\def\1{{\sst{(1)}}}
\def\2{{\sst{(2)}}}
\def\3{{\sst{(3)}}}
\def\4{{\sst{(4)}}}
\def\5{{\sst{(5)}}}
\def\6{{\sst{(6)}}}
\def\7{{\sst{(7)}}}
\def\8{{\sst{(8)}}}

 \def\bd{\begin{document}} \def\ed{\end{document}}
\def\ds{\documentstyle} \let\fr=\frac \let\bl=\bigl \let\br=\bigr
\let\Br=\Bigr \let\Bl=\Bigl
\let\bm=\bibitem
\let\na=\nabla
\let\pa=\partial \let\ov=\overline
\def\fft#1#2{{#1 \over #2}}

\newcommand{\insertplot}[5]{\begin{figure}
 \hfill\hbox to 0.05in{\vbox to #5in{\vfill
 \inputplot{#1}{#4}{#5}}\hfill}
 \hfill\vspace{-.1in}
 \caption{#2}\label{#3}
 \end{figure}}

    \numberwithin{equation}{section}

\begin{document}

\title{
Spinning black holes in Einstein--Gauss-Bonnet--dilaton theory:
non-perturbative solutions
}

\author{{\large Burkhard Kleihaus}$^{1}$, {\large  Jutta Kunz}$^{1}$,
  {\large Sindy Mojica }$^{1}$
	and
{\large Eugen Radu}$^{2}$ 
\\
\\
 {\small  $^{1}$Institut f\"ur Physik, Universit\"at Oldenburg, Postfach 2503
D-26111 Oldenburg, Germany}
\\
 {\small  $^{2}$Departamento de Fisica da Universidade de Aveiro and I3N,
 Campus de Santiago, 3810-183 Aveiro, Portugal} 
 }
\date{\today}

\maketitle

 \begin{abstract} 
We present an investigation of spinning black holes in Einstein--Gauss-Bonnet--dilaton (EGBd) theory.
The solutions are found
 within a non-perturbative approach, by directly solving
the field equations.
These stationary axially symmetric black holes are asymptotically flat. They
possess a non-trivial scalar field outside their regular event horizon.
We present an overview
of the parameter space of the solutions together with a study of their basic properties.
We point out that the EGBd black holes
can exhibit some physical differences when compared to the Kerr solution.
For example, their mass is always bounded from below, 
while their angular momentum  can exceed the Kerr bound,
Also, in contrast to the Kerr case, the extremal solutions are singular, 
with the scalar field diverging on the horizon.  
  \end{abstract}

\tableofcontents


\section{Introduction}

The Einstein--Gauss-Bonnet-dilaton (EGBd) gravity
is one of the most interesting and best motivated extensions of General Relativity (GR).
This model modifies the Einstein-Hilbert action by adding a (real) scalar field
non-minimally coupled to the Gauss-Bonnet (GB) invariant.
The resulting theory possesses a number of attractive features.
First, the EGBd model is the only theory of gravity with quadratic curvature
terms in the action leading to second order equations of motion.
Second, the terms in the EGBd action naturally occur in the framework
of the low-energy effective string theories, see $e.g.$ \cite{Moura:2006pz},
in which case the scalar field can be seen as the string dilaton.
Moreover, the EGBd gravity can  be considered as a
particular case of Horndeski gravity \cite{Kobayashi:2011nu},
a theory
which attracted recently a lot of interest.
A review of these aspects in a more general context 
can be found in the recent work 
\cite{Berti:2015itd}. 

Similar to the GR case, the EGBd model does not possess particle-like 
soliton solutions\footnote{Interestingly, as shown in 
\cite{Kanti:2011jz,Kanti:2011yv},
the EGBd model allows for traversable wormhole solutions, without needing any form of exotic matter.
}. 
However, there are black hole (BH) solutions, which 
intrinsically violate the `no hair' conjecture, possessing always a 
nonvanishing scalar field\footnote{In contrast to other models
 possessing solutions with scalar hair,
the (pure) Einstein gravity BHs do not satisfy the field equations 
of the full (EGBd) model.}. 
The study of BH solutions of the EGBd gravity
has started with the work \cite{Mignemi:1992nt,Mignemi:1993ce},  
where the generalization of the Schwarzschild solutions
was found in closed form within a perturbative approach.
The fully non-perturbative solution was constructed 
 numerically in
\cite{Kanti:1995vq,Torii:1996yi,Alexeev:1996vs,Guo:2008hf,Pani:2009wy}.
Perhaps the most interesting result there is the existence of
a minimal mass for the static BHs,
which is set by the  GB coupling constant.

In analogy to the GR case, these solutions should possess spinning 
generalizations.
However, their study is much more involved, 
and mainly perturbative results have been reported in the literature.
Here the most advanced results obtained so far are those in the recent work 
\cite{Maselli:2015tta}.
That paper
reports an approximate construction of the spinning solutions in EGBd theory
which is of fifth order in the black hole spin   and of seventh order in the
 GB coupling constant
(for previous work in this direction, see 
\cite{Pani:2009wy,Pani:2011gy,Ayzenberg:2014aka}).

The perturbative approach has the advantage to lead to closed form expressions 
of the solutions  and to  allow for a simple and direct computation of various 
quantities of interest.
At the same time, it is clear that a number of important features
(in particular those occurring in the fast spinning regime of the solutions)
cannot be caught within this framework. 

A different approach has been taken in Ref.~\cite{Kleihaus:2011tg}
which has  given a first discussion of the  
spinning solutions in EGBd theory within a 
non-perturbative framework. 
There, the field equations were solved numerically, and
the basic properties of the solutions were compared
with those of their GR counterparts.
Perhaps the most exciting result reported in 
\cite{Kleihaus:2011tg}
is  that the 
BHs in EGBd theory can slightly exceed the 
Kerr bound\footnote{
The specific angular momentum $a$ of Kerr BHs is bounded by
$|a| \le M$ with $a=J/M$.}.
Moreover, the innermost-stable-circular-orbits (ISCOs) can differ from the respective Kerr values.
Also, as discussed in \cite{Kleihaus:2014lba},
the quadrupole moment of EGBd black holes can be considerably larger than in the GR case.

The main purpose of this work is to provide a detailed description of the  
(non-perturbative) spinning BHs in EGBd theory reported in 
\cite{Kleihaus:2011tg}, with special emphasis 
on the construction of solutions together with their domain of existence.
The paper is organized
as follows.
In Section 2 we describe the EGBd theory, the Ansatz and the boundary conditions taken.
We also  introduce physical
quantities  and 
discuss the numerical procedure 
together with the known solutions. 
 This general framework is used in Section 3, where we present a discussion of the solutions.
Finally, in Section 4 we summarize our results and enumerate a number of research directions  
that can be addressed in the future.
In Appendix A we provide some details on
a set of  EGBd critical solutions which form a part of the boundary of the domain of existence,
while 
Appendix B contains a discussion of the issue of extremal solutions within the near horizon formalism.

\section{ The general framework}

\subsection{ The action  }
 
The Einstein--Gauss-Bonnet--dilaton (EGBd) system 
is described by the following action\footnote{In this work we shall use geometric units $c=G=1$.}
\begin{eqnarray}  
\label{act}
S=\frac{1}{16 \pi}\int d^4x \sqrt{-g} \left[R - \frac{1}{2}
 (\partial_\mu \phi)^2
 + \alpha e^{-\gamma \phi}R^2_{\rm GB}   \right],
\end{eqnarray} 
where 
$\phi$ is a (real) scalar field, and $\gamma$ and $\alpha$ are input parameters
of the theory. Also, 
\begin{eqnarray} 
\label{GB}
 R^2_{\rm GB} = R_{\mu\nu\rho\sigma} R^{\mu\nu\rho\sigma}
- 4 R_{\mu\nu} R^{\mu\nu} + R^2
\end{eqnarray} 
is the GB term\footnote{This term alone does not yield modifications of Einstein's equations,
its integral leading to a boundary term in the action.
This is no longer the case, however, once $R^2_{\rm GB}$ couples with dynamical matter fields.
 }. 
For the chosen conventions,
the (positive) GB coupling parameter\footnote{
In this work we shall treat $\alpha$ as an 
arbitrary input constant.
However, note that 
observations of BH low-mass X-ray binaries
give a constraint $\sqrt{\alpha} \lesssim 5\times 10^6$ cm \cite{Yagi:2012gp},
which
is comparable to the typical
size of a stellar-mass BH, see the discussion in \cite{Berti:2015itd}.
} $\alpha$ is dimensionful and, specifically,
it has dimensions of $(length)^2$.
Moreover, although all solutions reported in this work have $\gamma=1$,
we shall keep its value arbitrary in all general relations.

Varying the action~(\ref{act}) with respect to $g_{\mu\nu}$, we obtain
the  equations for the metric tensor
\begin{eqnarray}
\label{EGB-eq}
&& E_{\mu\nu}=G_{\mu\nu}
-\frac{1}{2} T_{\mu\nu}^{(\phi)}
+\alpha   e^{-\gamma\phi}\Bigl[H_{\mu\nu}
+4(\gamma^2\nabla^{\rho}\phi\nabla^{\sigma}\phi
-\gamma\nabla^{\rho}\nabla^{\sigma}\phi)P_{\mu\rho\nu\sigma}\Bigr]
=0,
\end{eqnarray}
where
\begin{eqnarray}
\nonumber
&&G_{\mu\nu}= R_{\mu\nu}-\frac{1}{2}g_{\mu\nu}R,~~~
T_{\mu\nu}^{(\phi)}=\nabla_{\mu}\phi\nabla_{\nu} \phi -\frac12 g_{\mu\nu}(\nabla\phi)^2,~~{\rm and}
\\
&&H_{\mu\nu}= 2\bigl(RR_{\mu\nu}-2R_{\mu \rho}R^{\rho}_{~\nu}
-2R^{\rho\sigma}R_{\mu\rho\nu\sigma}
+R_{\mu}^{~\rho\sigma\lambda}R_{\nu\rho\sigma\lambda}\bigr)
-\frac{1}{2}g_{\mu\nu}R^2_{\rm GB},
\\
\nonumber
&& P_{\mu\nu\rho\sigma}=
R_{\mu\nu\rho\sigma}+2g_{\mu[\sigma}R_{\rho]\nu}
+2g_{\nu[\rho}R_{\sigma]\mu} +Rg_{\mu[\rho}g_{\sigma]\nu}.
\label{EGB:eq}
\end{eqnarray}
In the above relations, $P_{\mu\nu\rho\sigma}$ is the divergence free part of the Riemann tensor,
$i.e.$
$\nabla_\mu P^{\mu}_{~\nu\rho\sigma}=0$.
One notes that formally, the eqs.~(\ref{EGB-eq}) can be written in an Einstein-like form as
\begin{eqnarray}
\label{EGB-eq1}
 G_{\mu\nu}=\frac{1}{2} T_{\mu\nu}^{\rm (eff)}~,
 \end{eqnarray} 
 with an effective energy-momentum tensor acquiring a contribution due to the
 GBd term
 \begin{eqnarray}
\label{Teff}
 T_{\mu\nu}^{\rm (eff)}=T_{\mu\nu}^{(\phi)}
-2\alpha  e^{-\gamma\phi}T_{\mu\nu}^{\rm (GBd)},
 \end{eqnarray} 
where
 \begin{eqnarray}
\label{TeffGB}
 T_{\mu\nu}^{\rm (GBd)}= H_{\mu\nu}
+4(\gamma^2\nabla^{\rho}\phi\nabla^{\sigma}\phi
-\gamma\nabla^{\rho}\nabla^{\sigma}\phi)P_{\mu\rho\nu\sigma} .
 \end{eqnarray}   
Variation of Eq.~(\ref{act}) with respect to the scalar field  leads
to a generalized Klein-Gordon equation, 
\begin{eqnarray}
\label{dil-eq}
\nabla^2 \phi -\alpha   \gamma e^{-\gamma\phi}  R^2_{\rm GB}
 =0.
\end{eqnarray}

 \subsection{ The Ansatz and equations of motion}
We are interested in stationary,
axially symmetric spacetimes possessing
two commuting Killing vector fields, $\xi$ and $\eta$, with
\begin{eqnarray}
\xi = \partial_t,~~~ {\rm and}~~~\eta=\partial_\varphi,
\end{eqnarray}
in a system of adapted coordinates.
Such spacetimes are usually described by a Lewis-Papapetrou--type Ansatz  \cite{Wald:rg},
which satisfies the circularity condition and contains four unknown functions.
In this work we shall use a version of this Ansatz as originally introduced in
\cite{Kleihaus:2000kg},
with a line element parametrization 
\begin{eqnarray}
\label{metric}
ds^2=- f dt^2 +  \frac{m}{f} \left( d r^2+ r^2d\theta^2 \right) 
           +  \frac{l}{f} r^2\sin^2\theta (d\varphi-\frac{\omega}{r} dt)^2 ,
\end{eqnarray}
with $r,\theta,\varphi$  
``quasi-isotropic" spherical coordinates and $t$ the time coordinate. Also,
$f$, $m$, $l$ and $\omega$ are functions of $r$ and $\theta$.
The scalar field is also a function of the coordinates $r$ and $\theta$ only, 
\begin{eqnarray}
\label{scalar}
\phi=\phi(r,\theta).
\end{eqnarray}

\subsection{Boundary conditions and asymptotic behaviour}

{\bf~~ Large $r$ asymptotics.}
The solutions in this work approach a Minkowski spacetime background as
$r\to \infty$. This implies the following boundary conditions\footnote{Setting $\phi|_{r=\infty}= 0$
removes the scaling symmetry of the equations $\phi\to \phi +c$, $r\to r e^{-\gamma c/2}$.}
\begin{eqnarray}
f|_{r=\infty}= m|_{r=\infty}= l|_{r=\infty}=1 \ , \ \ \
\omega|_{r=\infty}=\phi|_{r=\infty}= 0
\ . \label{bc1a} 
\end{eqnarray}
Since the scalar field is massless, 
one can construct an approximate solution of the field equations 
(\ref{EGB-eq}),
(\ref{dil-eq})
compatible with these asymptotics
as a power series in $1/r$.
The leading order terms on such an 
expansion are:  
\begin{eqnarray}
\nonumber
&&f=1-\frac{2 M}{r} +\frac{2 M^2}{r^2}
+\left(\frac{1}{3}M(C_1-4M^2)-2M_2 P_2(\cos\theta) \right)\frac{1}{r^3}
+ O \left(\frac{1}{r^4}\right),
\\
\label{inf-expr}
&&
m=1 + \frac{C_1}{r^2} - \frac{M^2+2 C_1+D^2/4}{r^2}\sin^2{\theta}
+ O\left( \frac{1}{r^3}\right),
\\
\nonumber
&&l=1+\frac{C_1}{r^2} + O\left(\frac{1}{r^3}\right),
~~~
\omega=-\frac{2 a M}{r^2}+\frac{6 a M^2}{r^3} + O\left(\frac{1}{r^4}\right),
~~\phi=-\frac{D}{r} +O\left(\frac{1}{r^3}\right),
 \end{eqnarray}
where $M$, $C_1$, $a$, $M_1$ and $D$ are free parameters, 
while $P_2(\cos \theta)$ is a Legendre polynomial of 2nd degree.
%

\bigskip
 
{\bf~~ Expansion on the event horizon.}
The event horizon of these stationary black hole solutions
resides at a surface of constant radial coordinate, $r=r_{\rm H}>0$,
and is characterized by the condition $f(r_{\rm H})=0$ \cite{Kleihaus:2000kg}.
At a regular horizon the metric functions must satisfy
\begin{eqnarray}
 \label{bc-horizon} 
f|_{r=r_{\rm H}}=
m|_{r=r_{\rm H}}=
l|_{r=r_{\rm H}}=0
\ , \ \ \ \omega|_{r=r_{\rm H}}=\omega_{\rm H},
\end{eqnarray}
where $\omega_{\rm H}$ is a constant, while the condition imposed on the scalar field is
\begin{eqnarray}
\partial_r \phi|_{r=r_{\rm H}}=0.
\end{eqnarray}
Again, it is possible to construct an approximate (power series) solution,
this time  
in terms of
\begin{eqnarray}
\delta=\frac{r}{r_{\rm H}}-1.
\end{eqnarray}
For non-extremal solutions (the case considered 
explicitly in this work),
the first terms in the near horizon expansion read
\begin{eqnarray}
&&
\label{horizon-expansion}
f(r,\theta)=\delta^2 f_2(\theta) (1 -\delta) + O(\delta^4) ,
~~
m(r,\theta)=\delta^2 m_2(\theta) (1 -3\delta) + O(\delta^4),
\\
&&
\nonumber
l(r,\theta)=\delta^2 l_2(\theta) (1 -3\delta) + O(\delta^4),
~~
\omega(r,\theta)=\omega_{\rm H} (1 + \delta) + O(\delta^2), 
~~
\phi(r,\theta)=\phi_{0}(\theta)  + O(\delta^2) ,
\end{eqnarray}
with $f_2$, $m_2$, $l_2$ and $\phi_0$ unspecified
functions\footnote{Note that, similar to the pure Einstein gravity case, 
the equation $E_{r}^\theta=0$ implies that the ratio $f_2^2/m_2$ is constant, a supplementary condition which is used as another 
test of the numerical accuracy. 
This also implies the constancy of the Hawking temperature, as given by (\ref{TH}).} and
$\omega_{\rm H}$ a constant.

\bigskip


{\bf~~Behaviour on the symmetry axis}
 
 The conditions of axial symmetry and regularity impose
the following boundary conditions on the symmetry axis,
$i.e.$ at $\theta=0,\pi$:
%
\begin{eqnarray}
\label{bc-axis}
& &\partial_\theta f|_{\theta={0,\pi}} =
   \partial_\theta m|_{\theta={0,\pi}} =
   \partial_\theta l|_{\theta={0,\pi}} =
   \partial_\theta \omega|_{\theta={0,\pi}} = 0 \ , 
\end{eqnarray}
for the metric function, 
while for the scalar field one imposes
\begin{eqnarray}
\partial_\theta \phi|_{\theta={0,\pi}} =0. 
\end{eqnarray}
Again, it is possible to construct an approximate form of the solutions near the symmetry axis
as a power series in $\theta$ (and $\pi-\theta$, respectively).
For example, the first terms in such an expansion as $\theta\to 0$ reads\footnote{ The
absence of conical singularities implies as well  
$m|_{\theta=0,\pi} =l|_{\theta=0,\pi}.$}
\begin{eqnarray}
&&
\nonumber
f(r,\theta)=\bar f_0(r)+\theta^2 \bar f_2(r) + O(\theta^4) ,
~
m(r,\theta)=\bar m_0(r)+\theta^2 \bar m_2(r) + O(\theta^4) ,
~
l(r,\theta)=\bar l_0(r)+\theta^2 \bar l_2(r) + O(\theta^4) ,
\\
&&
\label{z-expansion}
\omega(r,\theta)=\bar \omega_0(r)+\theta^2 \bar \omega_2(r) + O(\theta^4) ,
~~
\phi(r,\theta)=\bar \phi_0(r)+\theta^2 \bar \phi_2(r) + O(\theta^4) .
\end{eqnarray}

Also, all solutions discussed in this work are symmetric $w.r.t.$ 
a reflection on the equatorial plane, $\theta=\pi/2$.
Therefore, in the numerical calculations, it is sufficient to consider the range
$0\leq \theta \leq \pi/2$ for the angular variable $\theta$.
Then the metric functions and the scalar field are required to satisfy Neuman boundary conditions
in the equatorial plane, 
\begin{eqnarray}
\label{bc-equator}
\partial_\theta f|_{\theta=\pi/2} =
 \partial_\theta m|_{\theta=\pi/2} =
  \partial_\theta l|_{\theta=\pi/2} =
   \partial_\theta \omega|_{\theta=\pi/2} =
	\partial_\theta \phi|_{\theta=\pi/2} =0.
\end{eqnarray}

\subsection{General relations and quantities of interest}
Starting with the horizon properties, we note that
the solutions possess an event horizon of spherical topology,
the metric of a spatial cross-section of the horizon being 
\begin{eqnarray}
\label{horizon-metric}
d\Sigma^2=h_{ij} dx^i dx^j=r_{\rm H}^2\left(\frac{m_2(\theta)}{f_2(\theta)} d\theta^2+\frac{l_2(\theta)}{f_2(\theta)} \sin^2\theta d\varphi^2\right).
\end{eqnarray}
The Killing vector field
\begin{equation}
\chi = \partial_t - \frac{\omega_{\rm H}}{r_{\rm H}}\partial_{\varphi}
\  \label{chi} 
\end{equation}
is orthogonal to and null on the  horizon \cite{Wald:rg}.
The parameter $\omega_{\rm H} $ which enters the event horizon boundary conditions  (\ref{bc-horizon})
fixes the event horizon angular velocity $\Omega_{\rm H}$ of the BHs,
\begin{eqnarray}
\label{OmegaH}
\Omega_{\rm H}=-\frac{\xi^2}{\xi\cdot \eta}=-\frac{g_{\varphi t}}{g_{tt}}\bigg|_{r_{\rm H}}=\frac{\omega_{\rm H} }{r_{\rm H}}  .
\end{eqnarray}
As usual, we introduce the 
Hawking temperature $T_{\rm H}={\kappa}/({2\pi})$, where $\kappa$ is the surface gravity
defined as $\kappa^2=-\frac{1}{2}(\nabla_a \chi_b)(\nabla^a \chi^b)|_{r_{\rm H}}$,
which yields
 \begin{eqnarray}
\label{TH}
T_{\rm H}=\frac{1}{2 \pi r_{\rm H}}\frac{f_2(\theta)}{\sqrt{m_2(\theta)}} 
 \end{eqnarray}
(we recall $m_2(\theta),m_2(\theta)$ and $l_2(\theta)$ are functions which enter 
the near horizon expansion (\ref{horizon-expansion})).

The event horizon area of the BHs is given by 
 \begin{eqnarray}
\label{AH}
A_{\rm H}=2\pi r_{\rm H}^2\int_0^\pi d\theta \frac{\sqrt{l_2(\theta) m_2(\theta)}}{f_2(\theta)}.
 \end{eqnarray}
The Einstein gravity BHs
possess an entropy which is a quarter of the event horizon area.
However, the entropy of the EGBd BHs acquires an extra contribution which is induced by the GB term in the action.
Then the total entropy
can be written in Wald's form \cite{Wald:1993nt} 
as an integral over the event horizon
\begin{eqnarray}
\label{S-Noether} 
S=\frac{1}{4}\int_{\Sigma_{\rm H}} d^{2}x 
\sqrt{h}(1+ 2\alpha e^{-\gamma \phi} \tilde R),
\end{eqnarray} 
where $h$ is the determinant of the induced metric on the horizon 
(as given by (\ref{horizon-metric})),
and $\tilde R$ is the event horizon curvature. 
Its explicit form for the Ansatz in this work reads
\begin{eqnarray}
&&S=S_{\rm E}+S_{\rm GBd},~~{\rm with}~~S_{\rm E}= \frac{1}{2}\pi r_{\rm H}^2\int_0^\pi d\theta \frac{\sqrt{l_2 m_2}}{f_2},
~~~{\rm and}~~
\\
\nonumber
&&~~~S_{\rm GBd}= \frac{1}{2}\pi \alpha \int_0^\pi d\theta
\frac{ e^{-\gamma \phi_0}}{ l_2^{3/2}m_2^{5/2}} 
\bigg(
 2m_2^2l_2'^2 \sin\theta-m_2l_2(-3\sin\theta l_2' m_2'+4m_2(2\cos\theta l_2'+\sin\theta l_2''))
\\
\nonumber
&&{~~~~~~~~~~~~~~~~~}+l_2^2(8m_2^2\sin\theta-3\sin \theta m_2'^2+2m_2(3\cos \theta m_2'+\sin\theta m_2''))
\bigg)
,
\end{eqnarray}
with a prime denoting a derivative with respect to $\theta$.

Similar to GR solutions,
the total mass $M$ and the angular momentum $J$ are read from 
the asymptotic 
behaviour of the metric functions:
\begin{eqnarray}
\label{asym}
g_{tt} =-f+\frac{lw^2}{f}  \sin^2 \theta
=-1+\frac{2M}{r}+\dots,~~g_{\varphi t}=- \frac{l w r}{f} \sin^2 \theta=-\frac{2J}{r}\sin^2\theta+\nonumber \dots.  
\end{eqnarray}  
Moreover, the solutions possess a dilaton charge $D$, which is read 
from the first term in the far field asymptotics of the scalar field as given in (\ref{inf-expr}).

Remarkably,
as usual in (asymptotically flat, pure Einstein)  BH mechanics, 
the temperature, entropy and the global charges
are related through a Smarr mass formula, which reads
\begin{eqnarray}
\label{Smarr}
M=2 T_{\rm H} S+2\Omega_{\rm H} J-\frac{D}{2\gamma}.
\end{eqnarray} 
This Smarr relation is obtained by starting from the usual Komar-like expressions,
and making use of the equations of motion,
and the approximate form of the solution at the horizon, on the symmetry axis and at infinity\footnote{For example,
one uses the observation that   
both
$  E_\varphi^t \sqrt{-g}$ 
and
 $\left(R_t^t+ \alpha e^{-\gamma \phi}(T_t^{t{\rm(GBd)}}+\frac{1}{2}R_{\rm GB}^2) \right)\sqrt{-g}$ 
are total derivatives 
($i.e.$ they have expressions of the form $\partial_r U_r+\partial_\theta U_\theta$
with   $U_r$,  $U_\theta$
 complicated functions of ${\cal F}_i$ and their derivatives). 
Moreover,  $e^{-\gamma \phi} T_\mu^{\mu{\rm (GBd)}} \sqrt{-g}$
can also be written as a total derivative.
}.
The EGBd BHs 
satisfy also the first law
\begin{eqnarray}
\label{first-law}
dM=T_{\rm H} dS +\Omega_{\rm H} dJ .
\end{eqnarray} 

\subsection{The numerical approach}
The functions ${\cal F}_i=(f, m, l, \omega; \phi)$ $(i=1,\dots,4)$
are solutions of a complicated set of 
partial differential equations 
which are found by plugging the Ansatz 
(\ref{metric}), (\ref{scalar})
into the field eqs.~(\ref{EGB-eq}), (\ref{dil-eq}).
Starting with the generalized Klein-Gordon eq.~(\ref{dil-eq}),
we note that this reduces to 
a complicated relation involving first and second derivatives of all functions ${\cal F}_i$.
(This equation is too complicated to display here.)
Secondly, concerning the generalized Einstein equations (\ref{EGB-eq}) for the metric potentials, 
the only
non-trivial components are  
$E_t^t, E_r^r,E_\theta^\theta,E_\varphi^\varphi,E_\varphi^t$ and $E_r^\theta$.
Following the scheme originally proposed in \cite{Kleihaus:2000kg}
(for solutions of Einstein gravity coupled with non-Abelian fields),
 these six equations are divided into two groups.
First,  four of them are solved together with the scalar field
equation (\ref{dil-eq}).
This  yields a coupled system of five partial differential equations  on the five unknown functions ${\cal F}_i$
which is solved in practice. 
The four equations for the metric functions are a suitable linear combination of 
$E_t^t,E_\varphi^\varphi,E_\varphi^t$ and $E_r^r+E_\theta^\theta$
which diagonalizes\footnote{Note that, however, due to the contribution of the Gauss-Bonnet term, 
the final equations are not diagonal with respect to second derivatives
of ${\cal F}_i$.} the corresponding Einstein tensor parts
with respect to $\hat O f$, $\hat O l$,  $\hat O m$ and  $\hat O {\omega}$, respectively (with the 
  operator $\hat O=\partial_{rr}+\frac{1}{r^2}\partial_{\theta\theta}$).
The remaining two equations $E_r^r-E_\theta^\theta$
and $E_r^\theta$ are treated as constraints and used to check the accuracy of the numerical method.
Again, the explicit form of the generalized Einstein equations is too complicated to display here,
each equation containing around 340 independent terms.

Restricting the domain of integration to the region outside the horizon,
in the next step one introduces  a new radial variable  
$x=1-r_{\rm H}/r$ 
which maps the semi-infinite region $[r_{\rm H},\infty)$ to the closed region $[0,1]$.
This leads to the following substitutions in the differential equations
$r  {\cal F}_{,r}   \longrightarrow   \frac{1}{r_{\rm H}} (1-x) {\cal F}_{,x}$
and 
$
r^2 {\cal F}_{,rr}   \longrightarrow
\frac{1}{r_{\rm H}^2}
\big(
(1-x)^2  {\cal F}_{,xx}
  - 2 (1-x) {\cal F}_{,x} 
	\big)
	$
for each function ${\cal F}_i$.

The equations for ${\cal F}_i$ are then discretized on a non-equidistant grid in
$x$ and $\theta$. 
Typical grids used  have sizes $91 \times 51$,
covering the integration region
$0\leq x \leq 1$ and $0\leq \theta \leq \pi/2$.  
All numerical calculations  
are performed by using the professional package FIDISOL/CADSOL \cite{schoen},
which uses a  Newton-Raphson method.   
This code provides also an error estimate for each unknown function.
For the solutions in this work,
the typical  numerical error 
for the functions is estimated to be lower than $10^{-3}$. 
The Smarr relation (\ref{Smarr})  and the 1st law (\ref{first-law})
provide further tests of the numerical accuracy, leading to error estimates of the same order. 

In this approach, one provides three input parameters: 
$\alpha,~r_{\rm H}$ and $\Omega_{\rm H}=\frac{\omega_{\rm H}}{r_{\rm H}}$.
The  functions ${\cal F}_i$ are subject to the boundary conditions introduced in Section 2.
The quantities of interest are then computed from the numerical output.
(For example, the mass $M$,  and the angular momentum $J$
are extracted from the  asymptotic expressions (\ref{asym}),
while the Hawking temperature, the entropy and the horizon area 
are obtained from the event horizon data, etc.)

\subsection{Known limits of the model}
 
Before discussing the properties of the EGBd
spinning BHs, it is useful
to briefly review the properties of the solutions in two important limits
of the general model.

\subsubsection{Kerr metric: the $\alpha=0$ solutions}
The first case is found for $\alpha=0$, $i.e.$, an Einstein-(massless) scalar field theory.
Then, as implied by classical no-hair theorems (see $e.g.$ the recent review
\cite{Herdeiro:2015waa}), the scalar field vanishes, $\phi=0$,
and the only physically relevant spinning solution is the Kerr BH. 
This metric is given usually in Boyer-Lindquist coordinates;  
however, it can also be written by employing the metric Ansatz  (\ref{metric}) 
($i.e.$ a quasi-isotropic coordinate system),
with 
  \begin{eqnarray}
  \label{Kerr}
 f=\left(1-\frac{r_{\rm H}^2}{r^2}\right)^2\frac{F_1}{F_2},~~
 l=\left(1-\frac{r_{\rm H}^2}{r^2}\right)^2,
 ~~
 m=\left(1-\frac{r_{\rm H}^2}{r^2}\right)^2\frac{F_1^2}{F_2},~~
 \omega=\frac{2M\sqrt{M^2-4r_{\rm H}^2}}{r^2}\frac{(1+\frac{M}{r}+\frac{r_{\rm H}^2}{r^2})}{F_2},
 \end{eqnarray}
 where
\begin{eqnarray}
\nonumber
&&
F_1=\frac{2M^2}{r^2}+\left(1-\frac{r_{\rm H}^2}{r^2}\right)^2+\frac{2M}{r}\left(1+\frac{r_{\rm H}^2}{r^2}\right)-\frac{M^2-4r_{\rm H}^2}{r^2}\sin^2\theta,~~
\\
&&
\nonumber
F_2=\left(\frac{2M^2}{r^2}+\left(1-\frac{r_{\rm H}^2}{r^2}\right)^2+\frac{2M}{r}\left(1+\frac{r_{\rm H}^2}{r^2}\right)\right)^2
-\left(1-\frac{r_{\rm H}^2}{r^2}\right)^2\frac{M^2-4r_{\rm H}^2}{r^2}\sin^2\theta.
\end{eqnarray}
The above metric functions contain two parameters, $M$ -- the BH
mass,  and $r_{\rm H}$ -- the event horizon radius, with $M\geq 2 r_{\rm H}$.
To make contact with the numerical approach discussed above, one has to express $M$
as a function of 
 the horizon velocity (which enters the boundary conditions at $r=r_{\rm H}$).
This is done by inverting the relation
\begin{eqnarray}
 \Omega_{\rm H}=\frac{\sqrt{M^2 -4r_{\rm H}^2}}{2M(M+2r_{\rm H} )}. 
\end{eqnarray}

Then, one finds that given a fixed value of $\Omega_{\rm H}$, the Kerr BH
(written in quasi-isotropic coordinates)   
exhibits two branches of solutions.
 The first one starts with the Minkowski spacetime, 
and extends up
to a maximal value of $r_{\rm H}$,
where a second branch emerges and tends backwards towards
$r_{\rm H}=0$.
The maximal value of $r_{\rm H}$ depends on  $\Omega_{\rm H}$, with 
$r_{\rm H}^{(max)}=\frac{1}{2\Omega_{\rm H}}\frac{\sqrt{\frac{2}{1+\sqrt{5}}}}{3+\sqrt{5}} $.
As $r_{\rm H}\to 0$ on this second branch, an extremal BH is approached, which
saturates the Kerr bound, $J=M^2$.

The   entropy, angular momentum and the Hawking temperature of the Kerr solution are 
given by
\begin{eqnarray}
  S=2 \pi M(M+2 r_{\rm H}) , ~~J=M\sqrt{M^2-4 r_{\rm H}^2},~~
	T_{\rm H}=\frac{1}{4\pi M}\frac{1}{1+\frac{M}{2r_{\rm H}}},
\end{eqnarray}
 respectively.

\subsubsection{Spherically symmetric black holes in EGBd theory}
Another important limit is found by taking the static limit of the general model
with $\alpha \neq 0$.
The solutions in this case are spherically symmetric, 
and have been discussed in Refs.~\cite{Mignemi:1992nt}-\cite{Guo:2008hf}.
All these papers employ in their study
Schwarzschild-like coordinates.
However, the solutions can also be studied
by employing the metric Ansatz  (\ref{metric}),
with
$f(r,\theta)=f(r)$, $l(r,\theta)=m(r,\theta)=m(r)$, $\omega(r,\theta)=0$
and $\phi(r,\theta)=\phi(r)$.
Thus the corresponding  Ansatz in isotropic coordinates reads
 \begin{eqnarray}
ds^2=\frac{m(r)}{f(r)}\left(dr^2+r^2(d\theta^2+\sin^2 \theta d\varphi^2)\right)-f(r) dt^2~~ {\rm and }~~\phi=\phi(r).
 \end{eqnarray}
The metric functions $f(r)$, $g(r)$ and the scalar field $\phi(r)$ can be found by solving
a system of ordinary differential equations.

The solutions possess a relatively simple
 near horizon expansion with (we recall $\delta=r/r_{\rm H}-1$)
  \begin{eqnarray}
f(r)=f_2\delta^2(1-\delta)+\dots,~~m(r)=m_2\delta^2(1-3\delta)+\dots,~~
\phi(r)=\phi_0+\phi_2\delta^2+\dots,
 \end{eqnarray}
with $f_2$, $m_2$ and $\phi_0$
parameters fixed by the numerical calculations, and
\begin{eqnarray}
\label{phi2}
\phi_2=\frac{e^{3\gamma\phi_0}m_2^3r_{\rm H}^4}{768 \alpha^3f_2^3\gamma^3}
\left(
1- \sqrt{1-\frac{96 e^{-2\gamma\phi_0}\alpha^2f_2\gamma^2  }{m_2^2r_{\rm H}^4}}
\right)^2,
\end{eqnarray}
while the corresponding expressions for $r\to \infty $ can be read from (\ref{inf-expr}).

An interesting feature of these BHs is the existence, for a given $\alpha$, of a minimal horizon size of the
 solutions \cite{Kanti:1995vq}.
This follows from the `reality' condition as imposed by the expression (\ref{phi2}) for $\phi_2$,
\begin{eqnarray}
 \label{cond}
 1-\frac{96 e^{-2\gamma\phi_0}\alpha^2f_2^2\gamma^2  }{m_2^2r_{\rm H}^4}>0,
 \end{eqnarray}
which can also be written as
  \begin{eqnarray}
 \label{cond1}
 {A_{\rm H}}    >{16\pi\sqrt{6} \alpha \gamma e^{-\gamma\phi_0} }.
 \end{eqnarray} 
At $r_{\rm H}^{(cr)}$ the solution becomes singular
\cite{Torii:1996yi}-\cite{Guo:2008hf}. 
 
A close inspection of the solutions and their properties reveals
that 
very near to the critical minimal value of the horizon radius
$r_{\rm H}^{(cr)}$ a minuscule second branch of BH solutions can arise
\cite{Torii:1996yi}-\cite{Guo:2008hf}.
Along this branch the mass then increases with decreasing horizon radius.
For $\gamma=1$ it covers the range 
\begin{eqnarray}
\label{condj}
2.4053 \leq\frac{M}{\sqrt{\alpha}}\lesssim 2.4055~.
 \end{eqnarray} 
The existence of this minuscule second branch depends on the value of $\gamma$
\cite{Guo:2008hf}, $e.g.$, for $\gamma=1/2$ is it no longer present.

Consequently, as discussed in \cite{Kanti:1995vq}-\cite{Pani:2009wy},
a lower bound is imposed on the BH mass,
which for $\gamma =1$ reads\footnote{Note, that for $\gamma=1$
the minuscule second branch 
exists only for
$\displaystyle 0.17284 \leq\frac{\alpha}{M^2}\lesssim 0.17285$
and can thus be neglected unless high accuracy is demanded.}
  \begin{eqnarray}
 \label{cond2}
0\leq\frac{\alpha}{M^2}\lesssim 0.1728~.
 \end{eqnarray} 
Then for a given mass, the static EGBd BHs with $\gamma=1$ exist for a limited range of 
the entropy, of the area and of the temperature, with   
	  \begin{eqnarray}
 \label{cond3}
4\pi \leq\frac{S}{M^2}\lesssim 15.082,~~
16\pi \leq\frac{A_{\rm H}}{M^2}\lesssim 42.842~~{\rm and}~~
\frac{1}{8\pi}\leq T_{\rm H} M \lesssim  0.043.
 \end{eqnarray} 

 Finally let us mention that the scalar charge $D$ is not an independent quantity,
but depends on the BH mass. Thus the scalar hair is of secondary type.

\section{The  properties of EGBd spinning black holes 
}

\subsection{Constructing the solutions and limits}

\begin{figure}[h!]
\begin{center}
\includegraphics[height=.2225\textheight, angle =0]{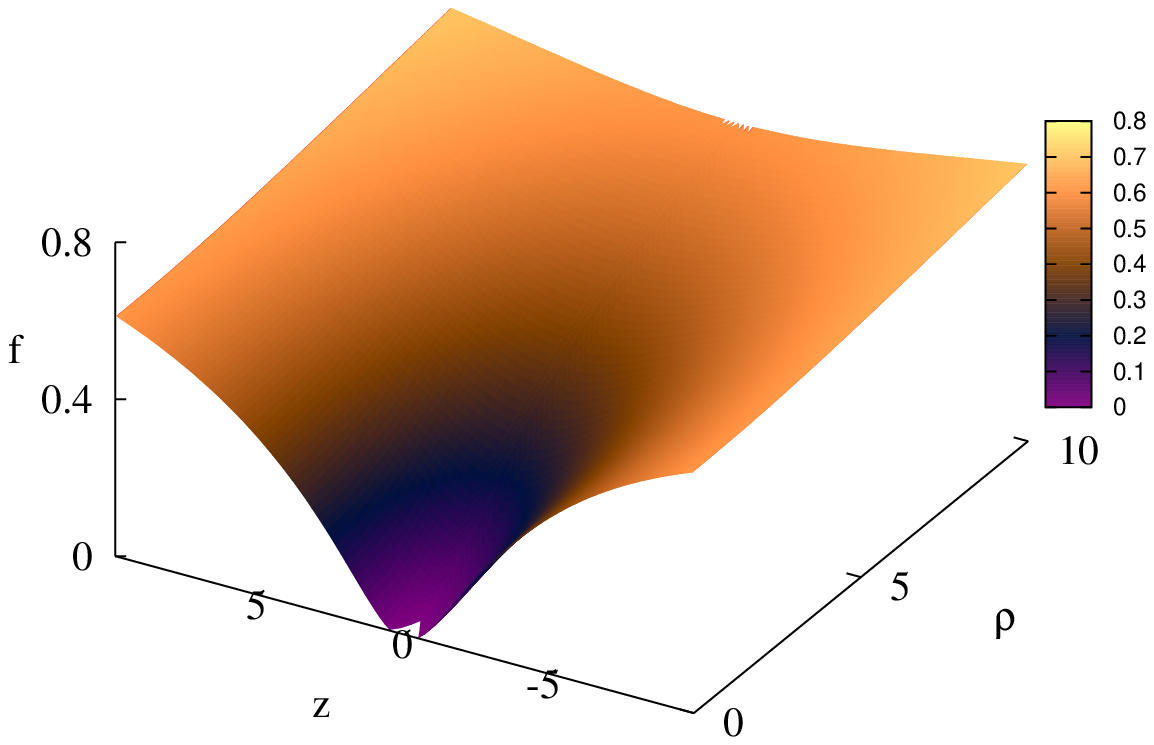} 
\includegraphics[height=.2225\textheight, angle =0]{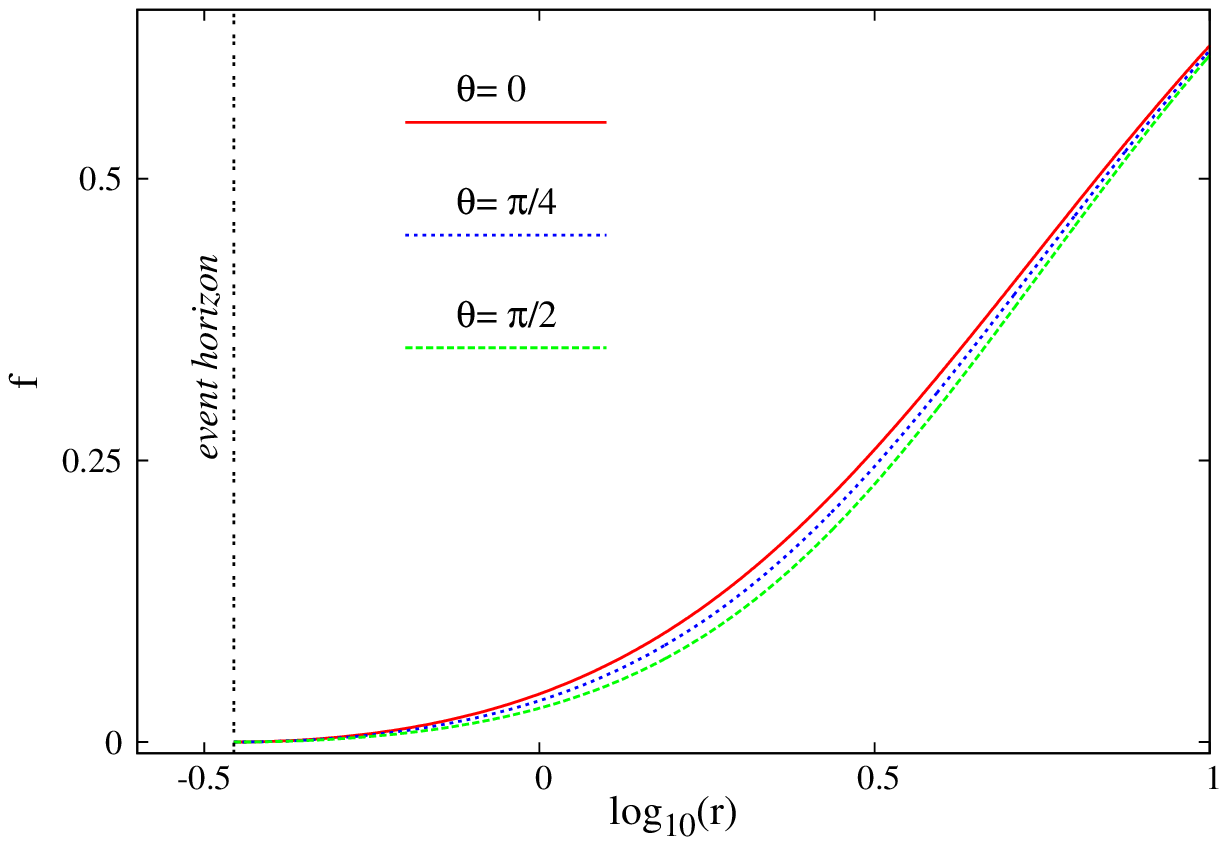} \ \ 
\includegraphics[height=.2225\textheight, angle =0]{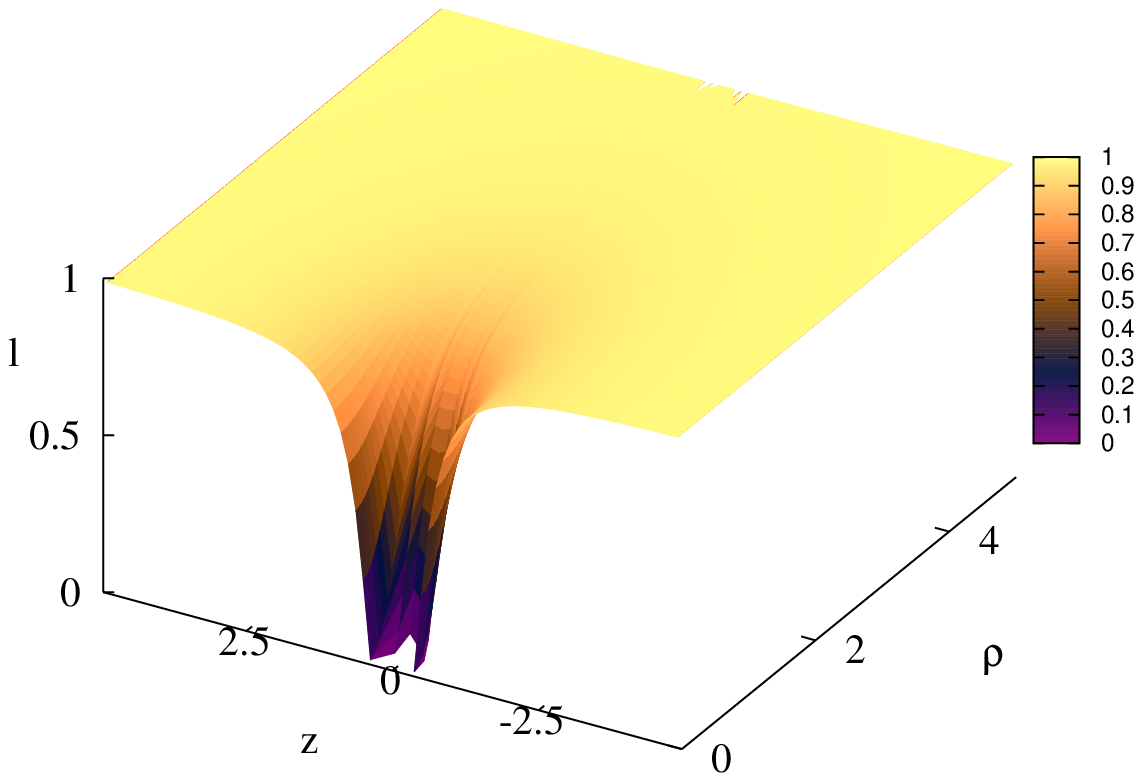} 
\includegraphics[height=.2225\textheight, angle =0]{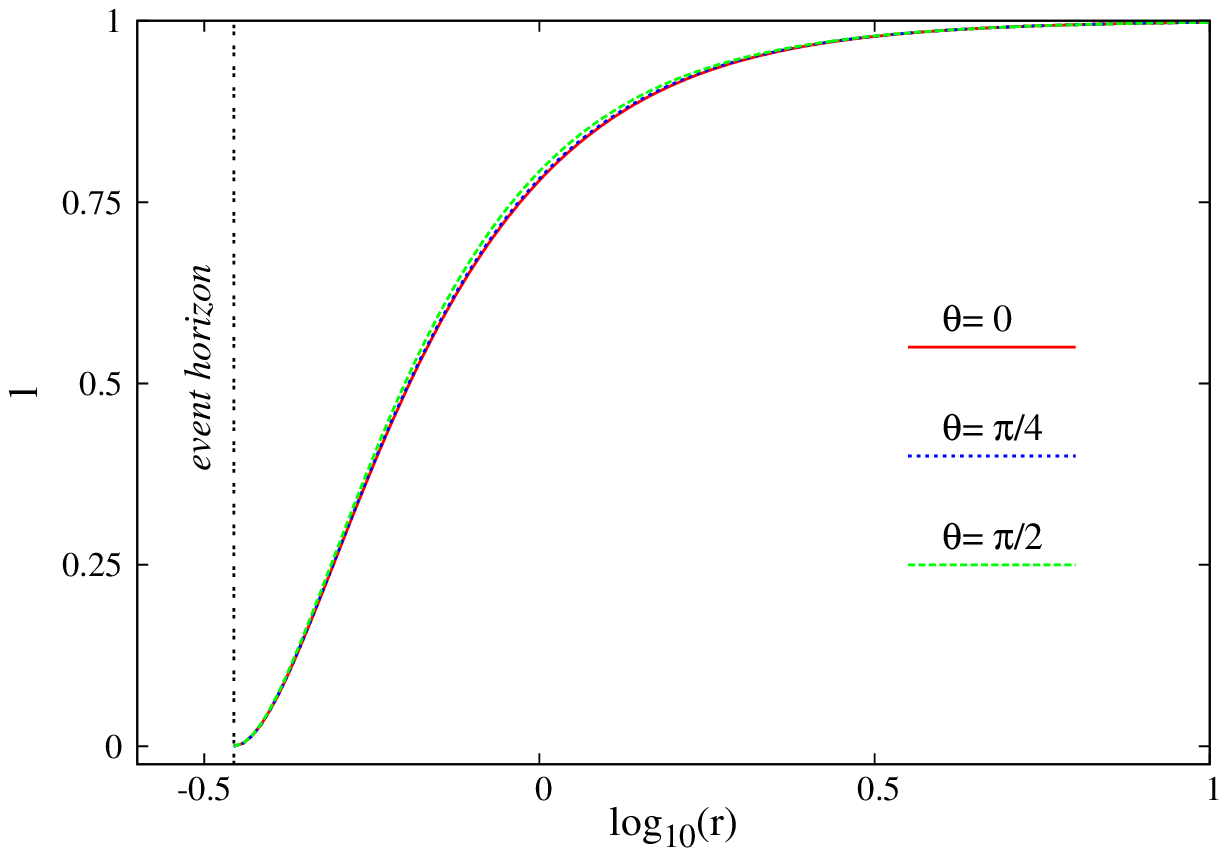} \ \
\includegraphics[height=.2225\textheight, angle =0]{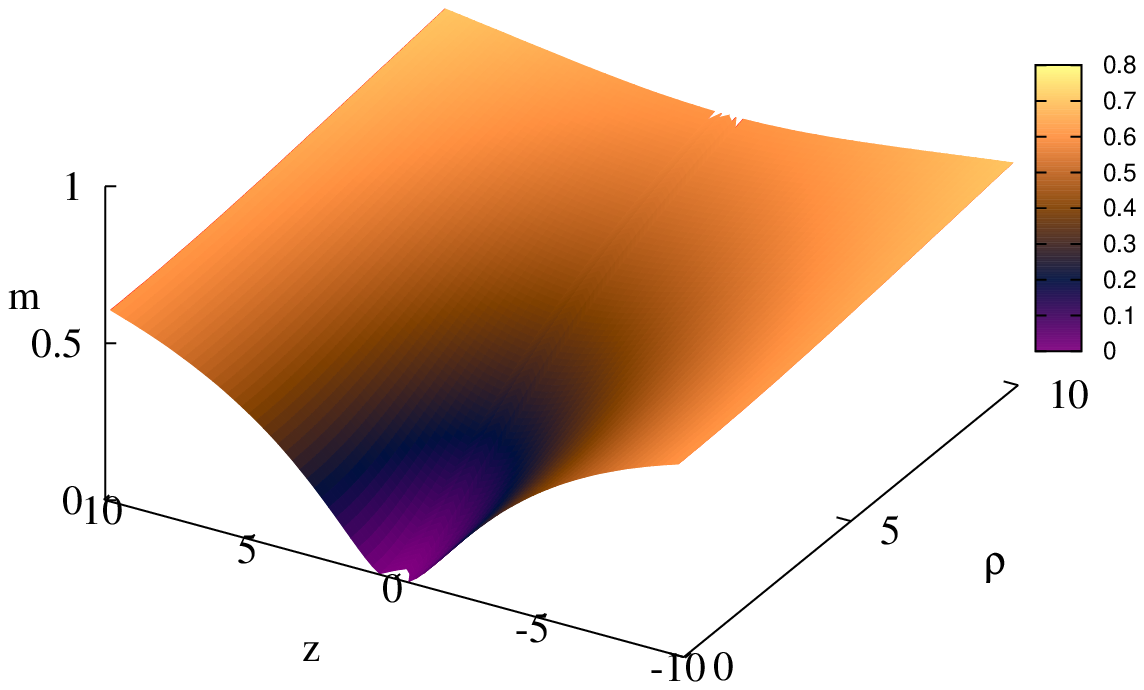} 
\includegraphics[height=.2225\textheight, angle =0]{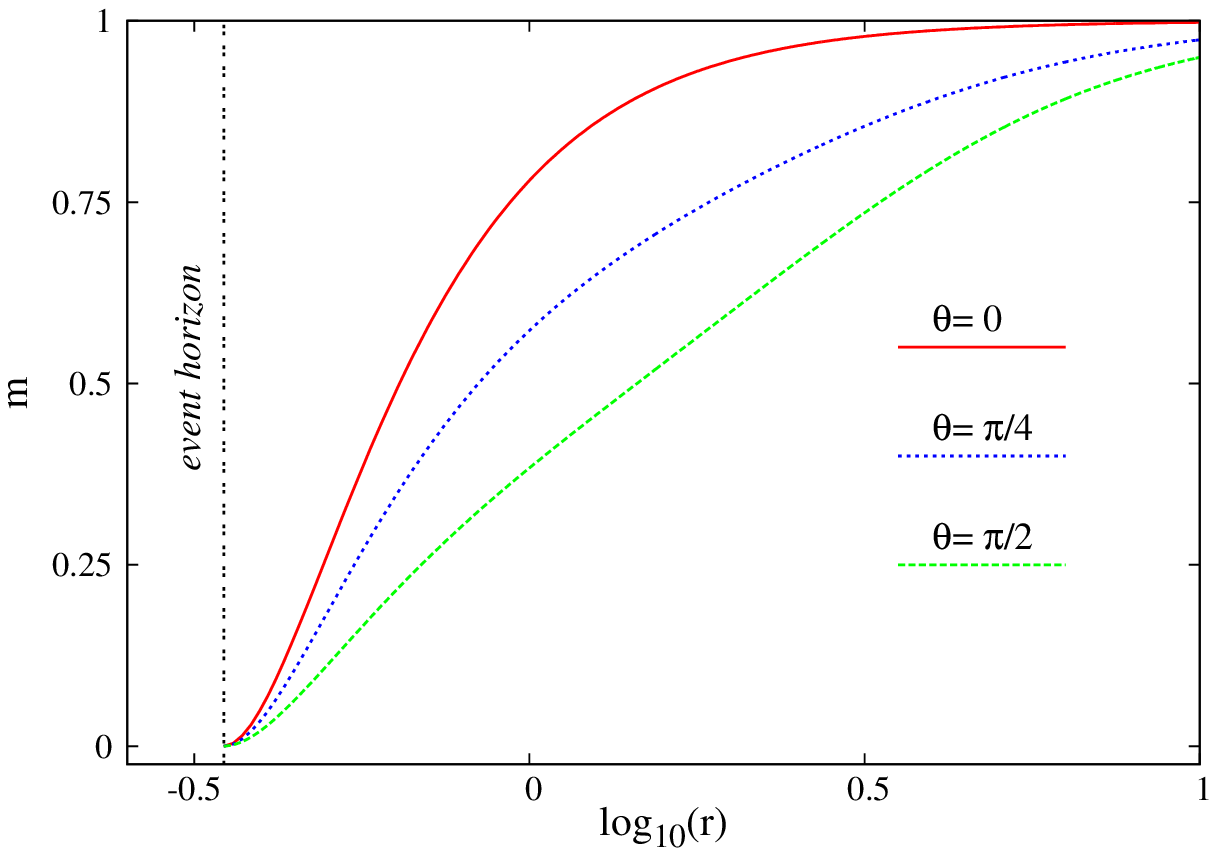} \ \  
\includegraphics[height=.2225\textheight, angle =0]{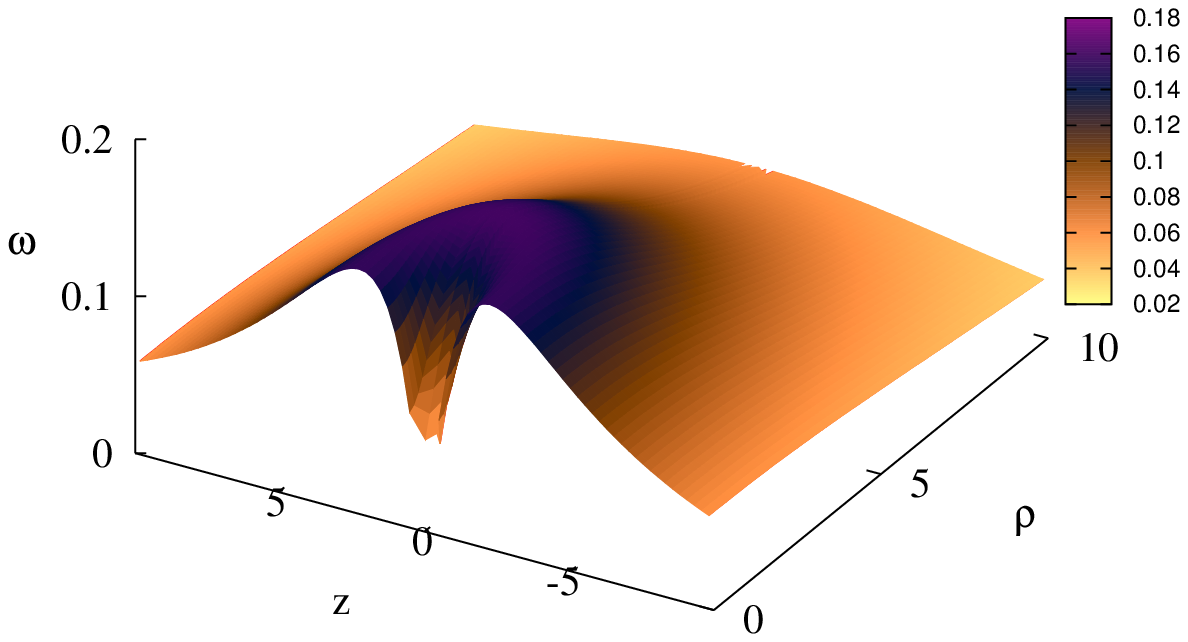} 
\includegraphics[height=.2225\textheight, angle =0]{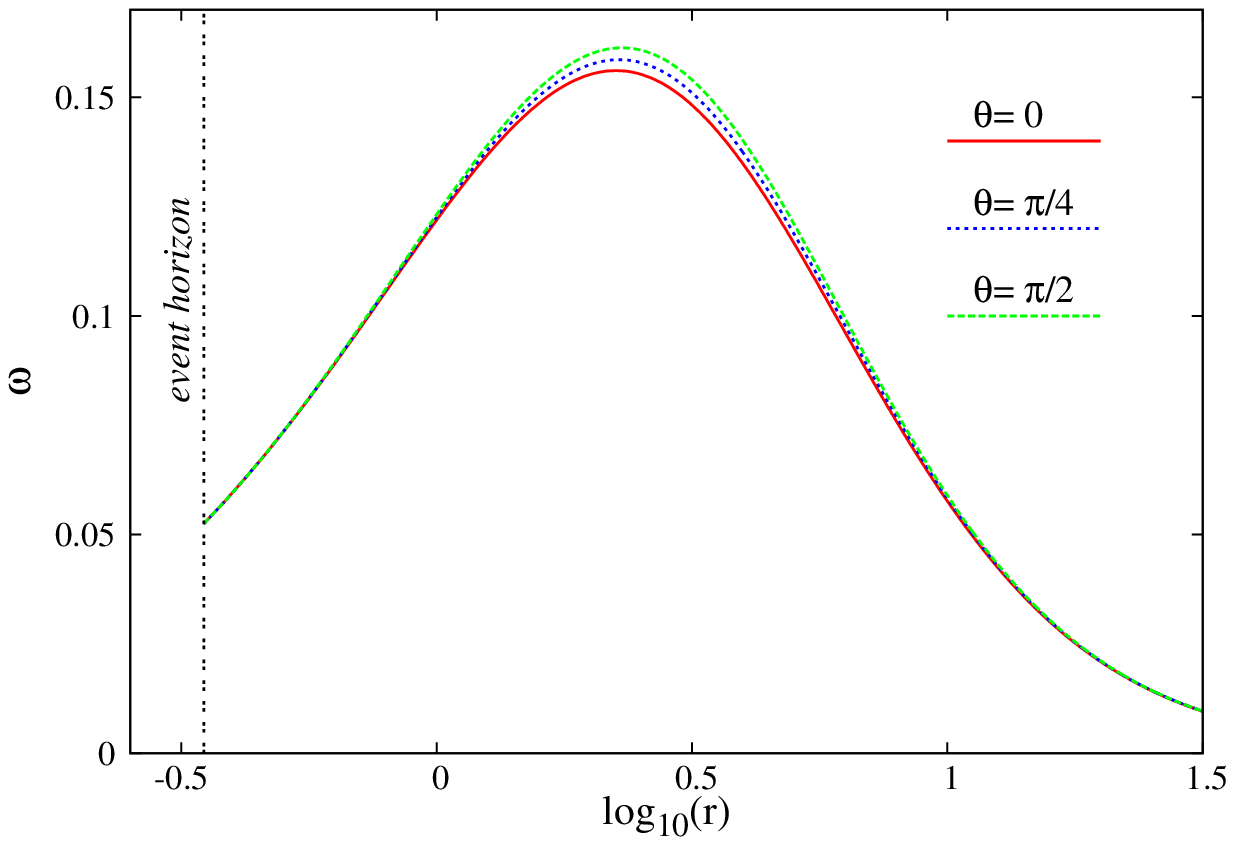}  
\end{center}
  \vspace{-0.5cm}
\caption{The metric functions $f,l,m$ and $\omega$ are shown as for a typical
EGBd spinning black hole with the input parameters $r_{\rm H}=0.35$, $\Omega_{\rm H}=0.15$ and $\alpha=0.5$.
This solution has $M=2.472$, $J=5.886$, $D=0.231$, $T_{\rm H}=.00737$, $A_{\rm H}=193.401$ and $S=55.751$.
}
\label{sol1}
\end{figure}

\begin{figure}[h!]
\begin{center}
\includegraphics[height=.26\textheight, angle =0]{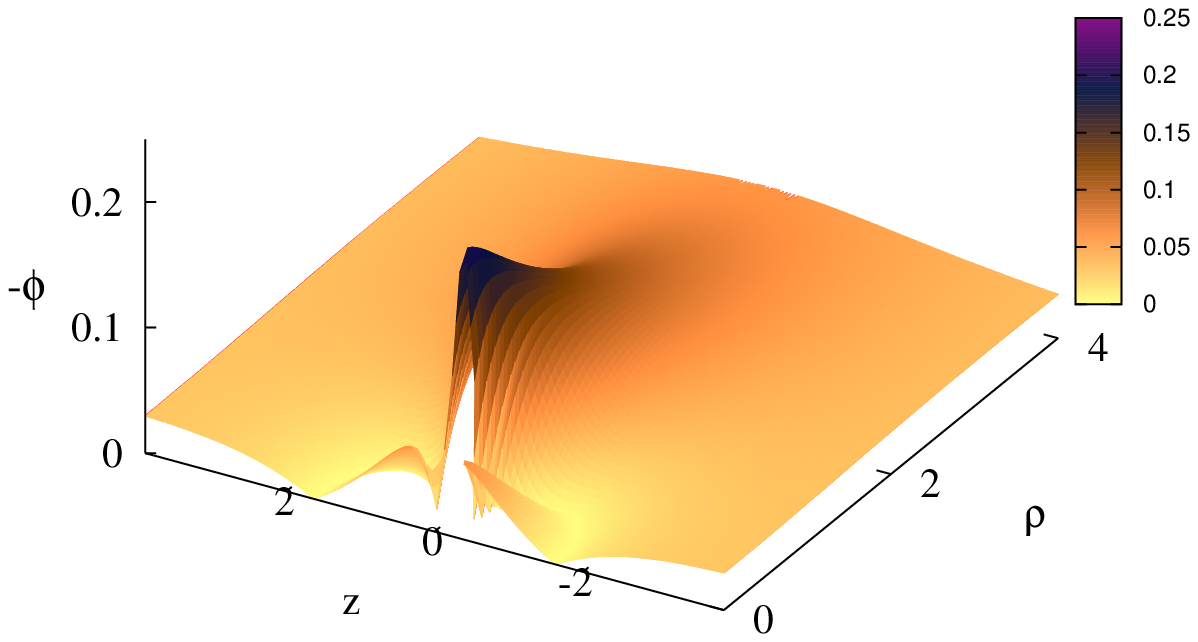} 
\includegraphics[height=.26\textheight, angle =0]{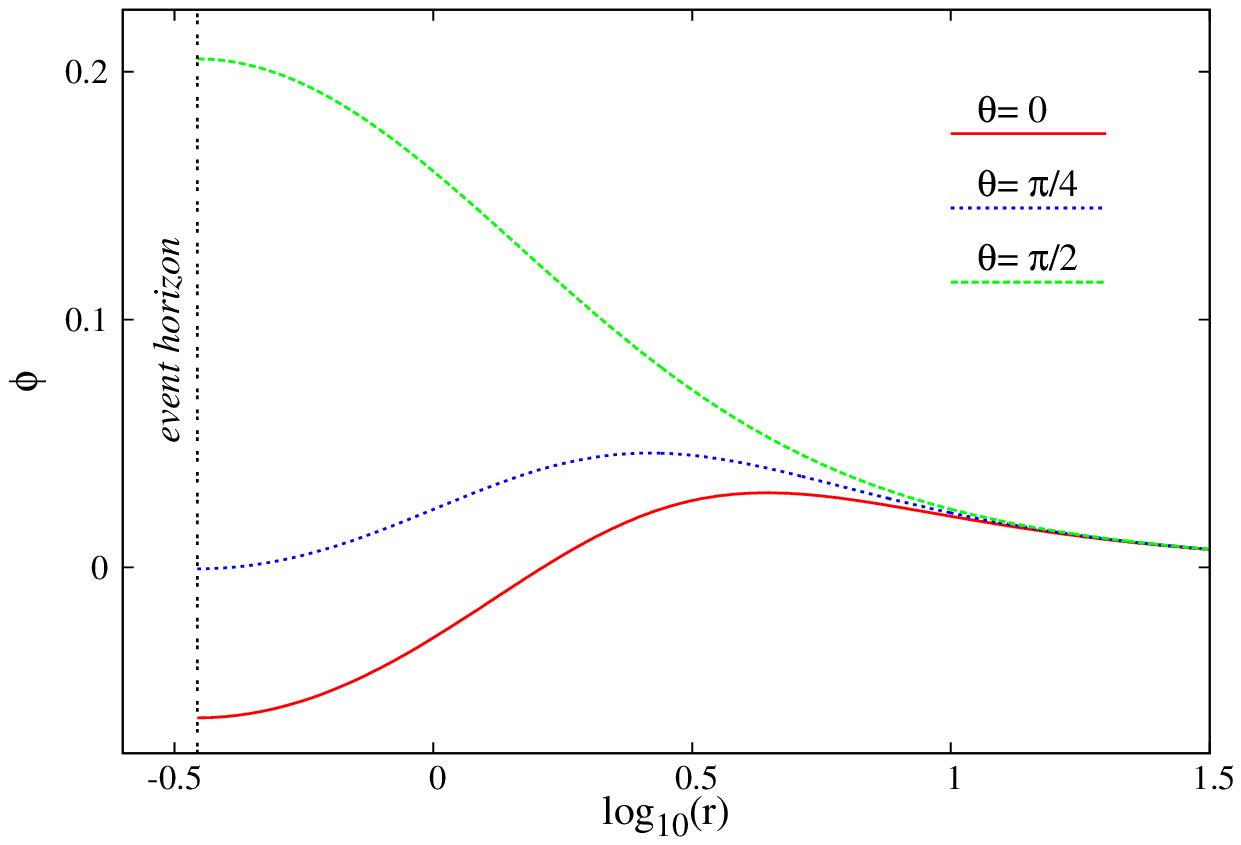} \ \
\includegraphics[height=.26\textheight, angle =0]{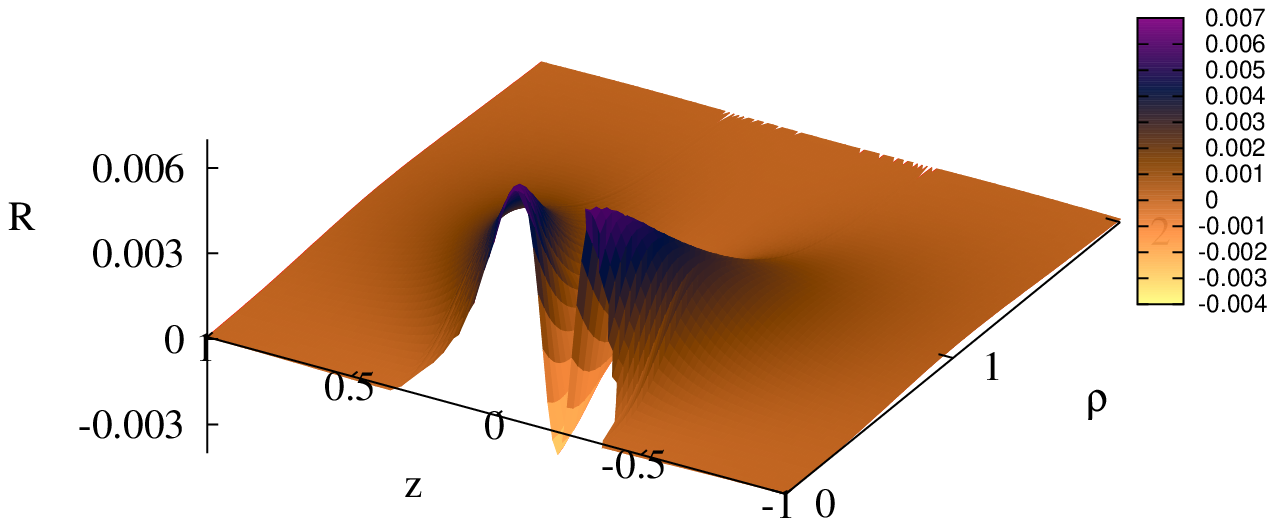} 
\includegraphics[height=.26\textheight, angle =0]{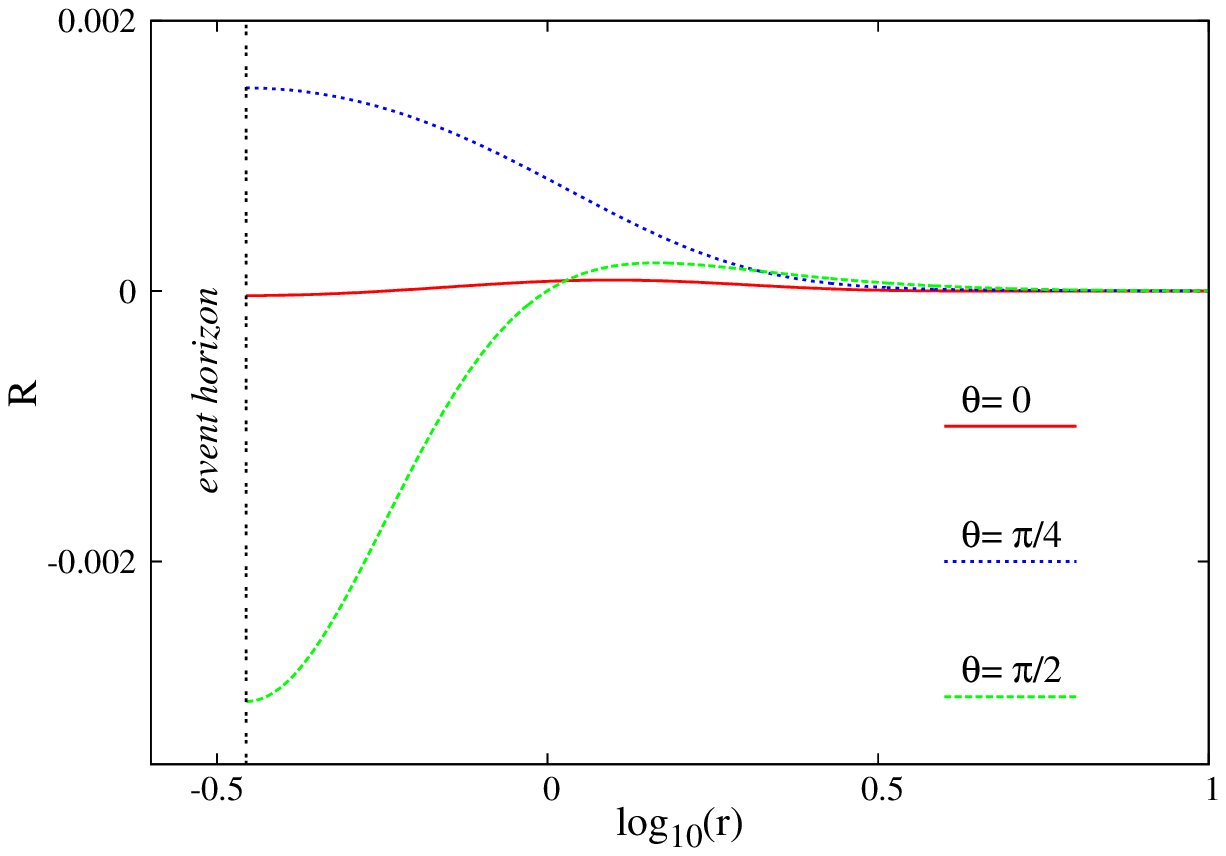} \ \ 
\includegraphics[height=.26\textheight, angle =0]{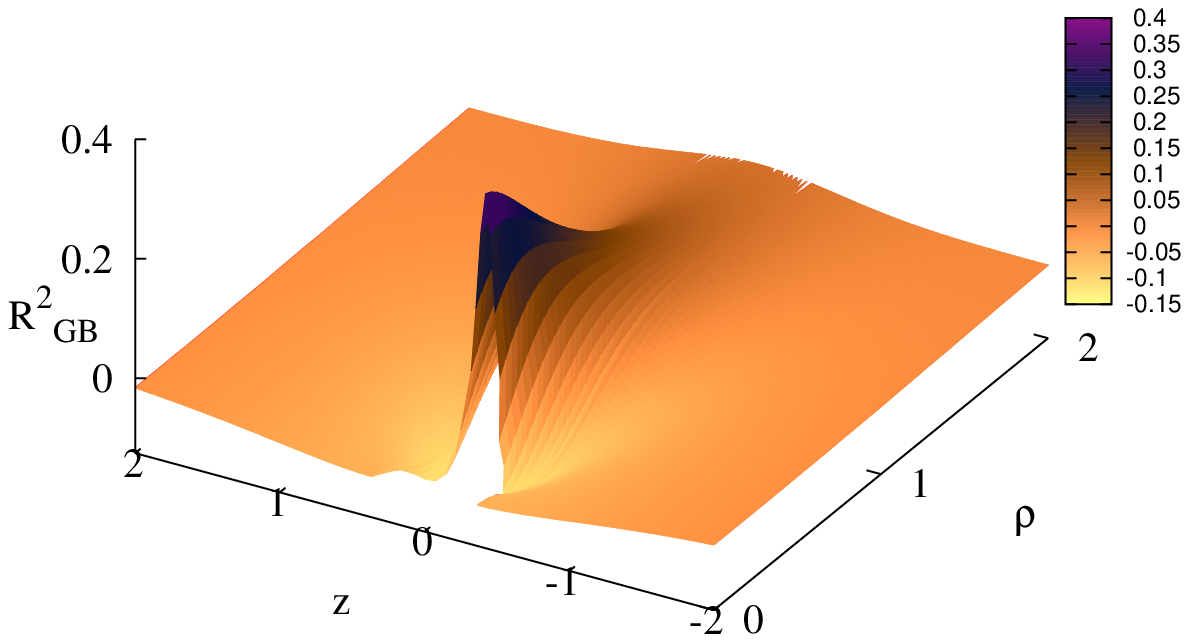} 
\includegraphics[height=.26\textheight, angle =0]{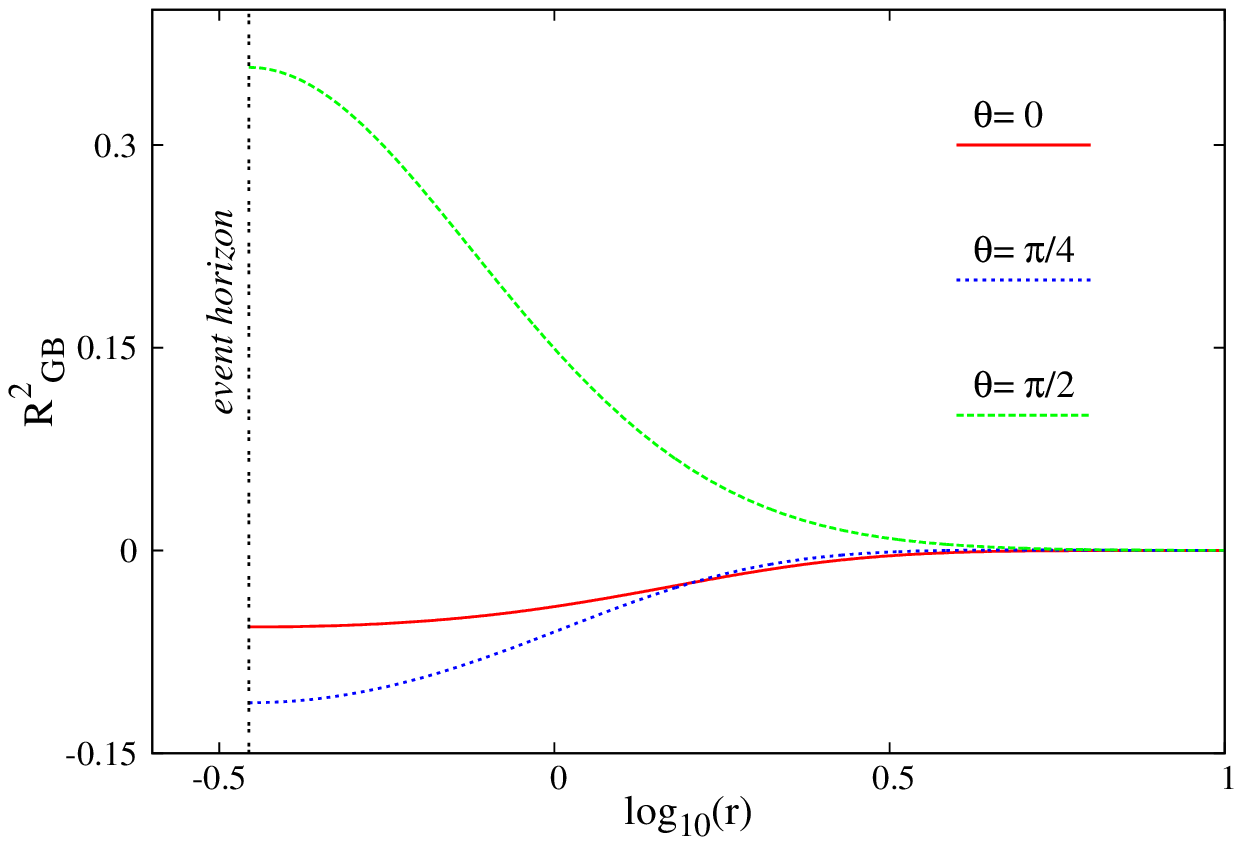}  
\end{center}
  \vspace{-0.5cm}
\caption{ The scalar field $\phi$ 
is shown 
together with the scalar curvature $R$ and 
the Gauss-Bonnet term (\ref{GB}) for the same solution as in Figure \ref{sol1}. }
\label{sol2}
\end{figure}

\begin{figure}[h!]
\begin{center}  
\includegraphics[height=.26\textheight, angle =0]{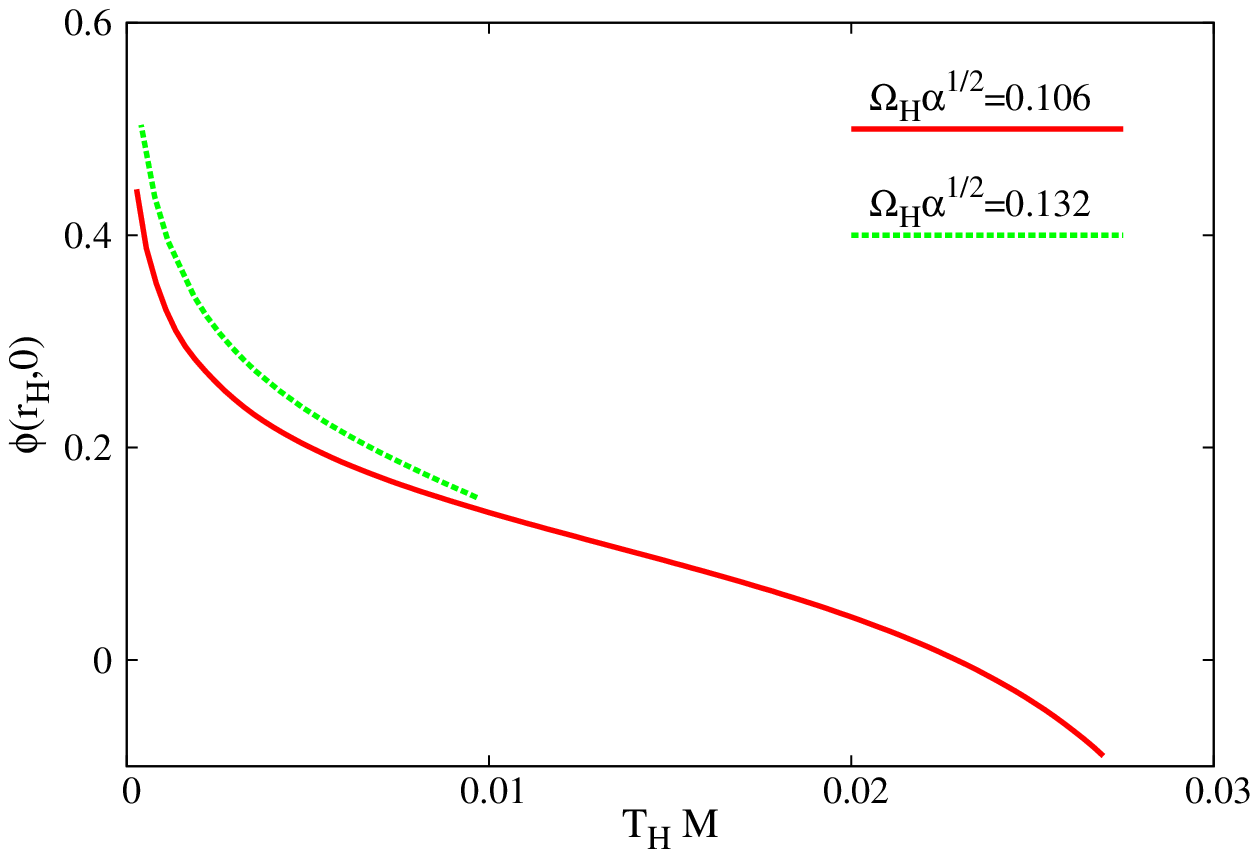} 
\end{center}
  \vspace{-0.5cm}
\caption{The value of the scalar field at the poles of the horizon
is shown as a function of the scaled temperature for two values of 
the scaled angular velocity $\Omega_{\rm H}\sqrt{\alpha}$.}
\label{ZH}
\end{figure}

We start by recalling that all solutions 
in this work have a fixed value  $\gamma=1$ of the dilaton coupling constant.

As expected, all spherically symmetric EGBd BHs discussed above possess rotating generalizations.
In principle, they can be constructed in closed form by considering a (double) perturbative approach in terms of the 
dimensionless parameters
$\alpha/M^2$ and $J/M^2$, see $e.g.$ \cite{Maselli:2015tta}.

The nonperturbative solutions are found by directly solving 
the EGBd equations for the functions ${\cal F}_i$
without any approximation.  
For all solutions we have found,
the metric functions $\mathcal F_i$ (and their first and second derivatives with respect to $r$ and $\theta$) have 
smooth profiles, which leads to finite curvature invariants on the full domain of integration, in particular
on the event horizon.
The shape of the functions $f,l,m$ and $\omega$ is rather similar to the GR case. 

To illustrate these features, we display in Figures 
\ref{sol1} and \ref{sol2}
the profile functions of a typical solution 
together with the Ricci and the Gauss-Bonnet scalars.
There, the left column shows 3D plots, whereas the
right column displays 2D plots of the corresponding functions in terms of the radial variable for three different values of the
angular coordinate;
the axes for the 3D plots are $\rho = r \sin\theta $ and
$z= r \cos\theta $  (with $r\geq  r_{\rm H}$). 
For example, from  Figure  \ref{sol2} (top panel)
one can see that the scalar field possesses 
a rather complicated shape, with a nontrivial dependence on the
angular coordinate $\theta$. 
Also, as seen in Figure  \ref{sol2}, both $R$ and $R_{\rm GB}^2$
stay finite everywhere, in particular at the horizon.

In our approach,
the solutions are computed for fixed sets of $\Omega_{\rm H}$ and $\alpha$,
while the (quasi-isotropic) horizon radius $r_{\rm H}$ 
and consequently the event horizon area $A_{\rm H}$
is varied\footnote{In principle,
it is sufficient to consider a single value of $\Omega_{\rm H}$ and scan the full range of $\alpha$. 
In practice, however, some regions of the parameter space are explored more 
easily by varying both 
 $\Omega_{\rm H}$ and $\alpha$.
}.  
As in the Kerr case, discussed above,
we typically find two branches of solutions\footnote{However, 
for sufficiently large values of $\Omega_{\rm H} \sqrt{\alpha}$
 (e.g., $\Omega_{\rm H}\sqrt{\alpha}=0.11$) one finds a single branch of solutions, 
with $0<r_{\rm H}<r_{\rm H}^{(cr)}$.}   in terms of the quasi-isotropic 
horizon radius $r_{\rm H}$.
These two branches then merge and end at a maximal value of $r_{\rm H}$ 
which depends on  $\Omega_{\rm H}$ and $\alpha$.
However, whereas in the Kerr case, the first branch starts
with the Minkowski spacetime, in the EGBd case the horizon area
cannot shrink to zero.
Indeed, similar to the static case, 
the rotating BH solutions on the first branch 
cease to exist for $r_{\rm H}$ smaller
than some critical value $r_{\rm H}^{(cr)}(\alpha,\Omega_{\rm H})$.
We note, that as $r_{\rm H}\to r_{\rm H}^{(cr)}$,
the numerical 
integration becomes delicate.
The `technical' reason which causes the solutions to
cease to exist beyond $r_{\rm H}^{(cr)}$ corresponds to
a more complicated version of the one found in the static case.
It is discussed in Appendix A,
and again based on the analysis of
the field equations at the event horizon.

\newpage

\begin{figure}[p!]
\begin{center}

\includegraphics[height=.225\textheight, angle =0]{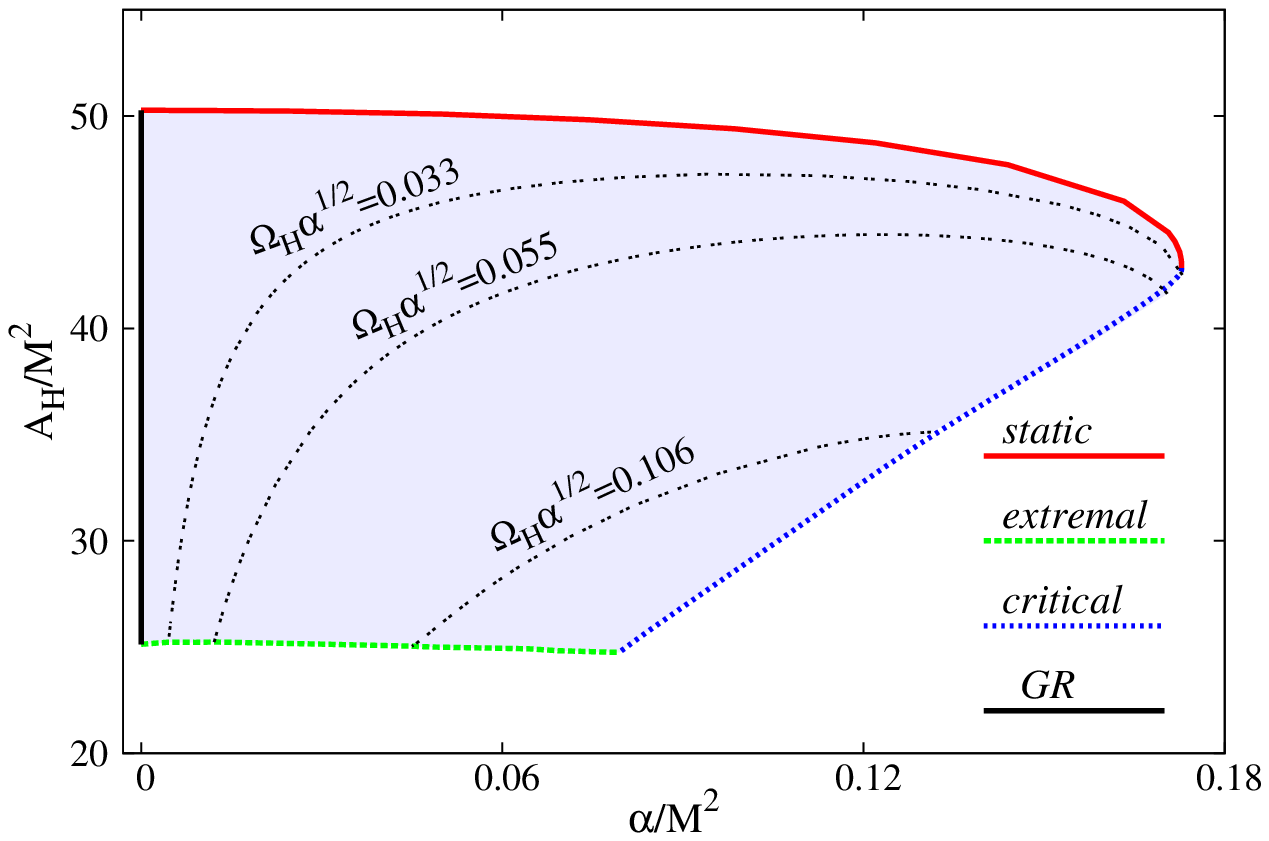} 
\includegraphics[height=.225\textheight, angle =0]{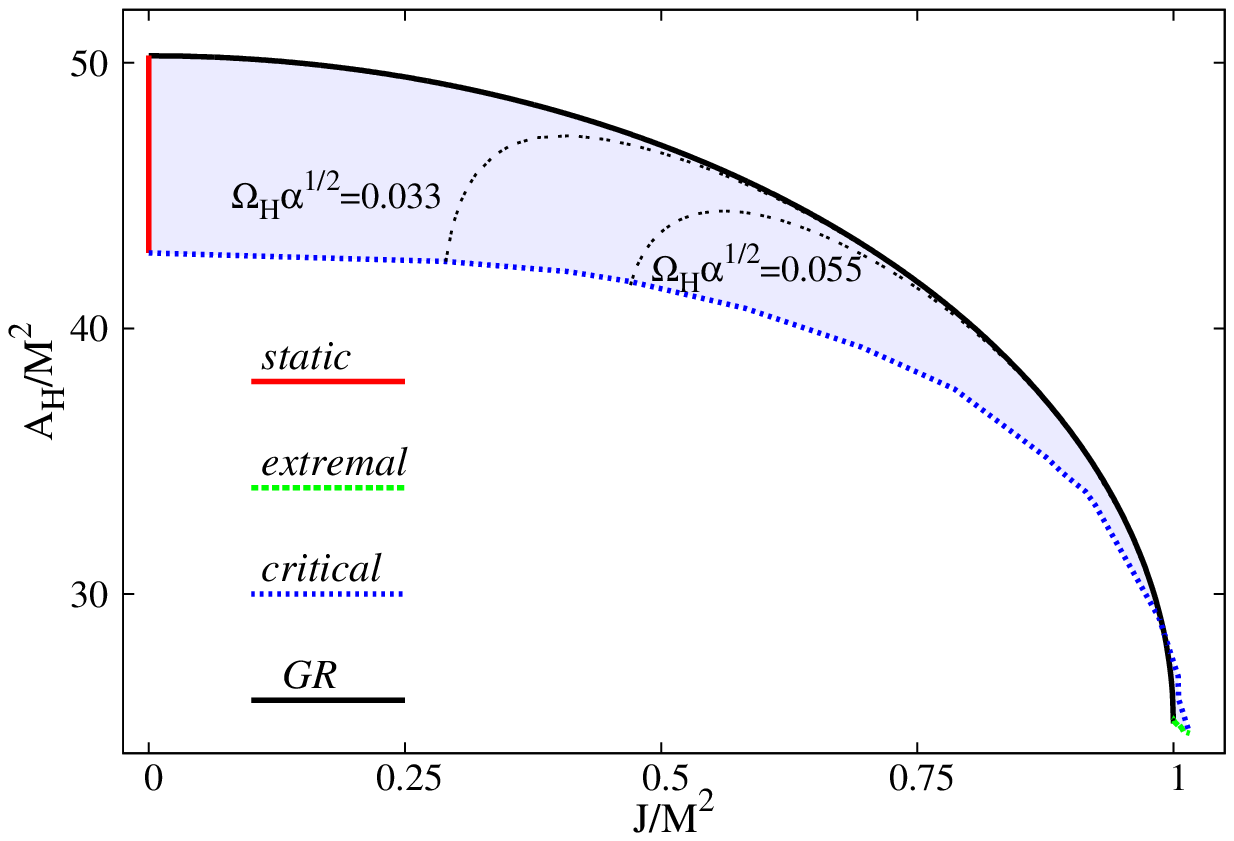} \ \ 
\includegraphics[height=.225\textheight, angle =0]{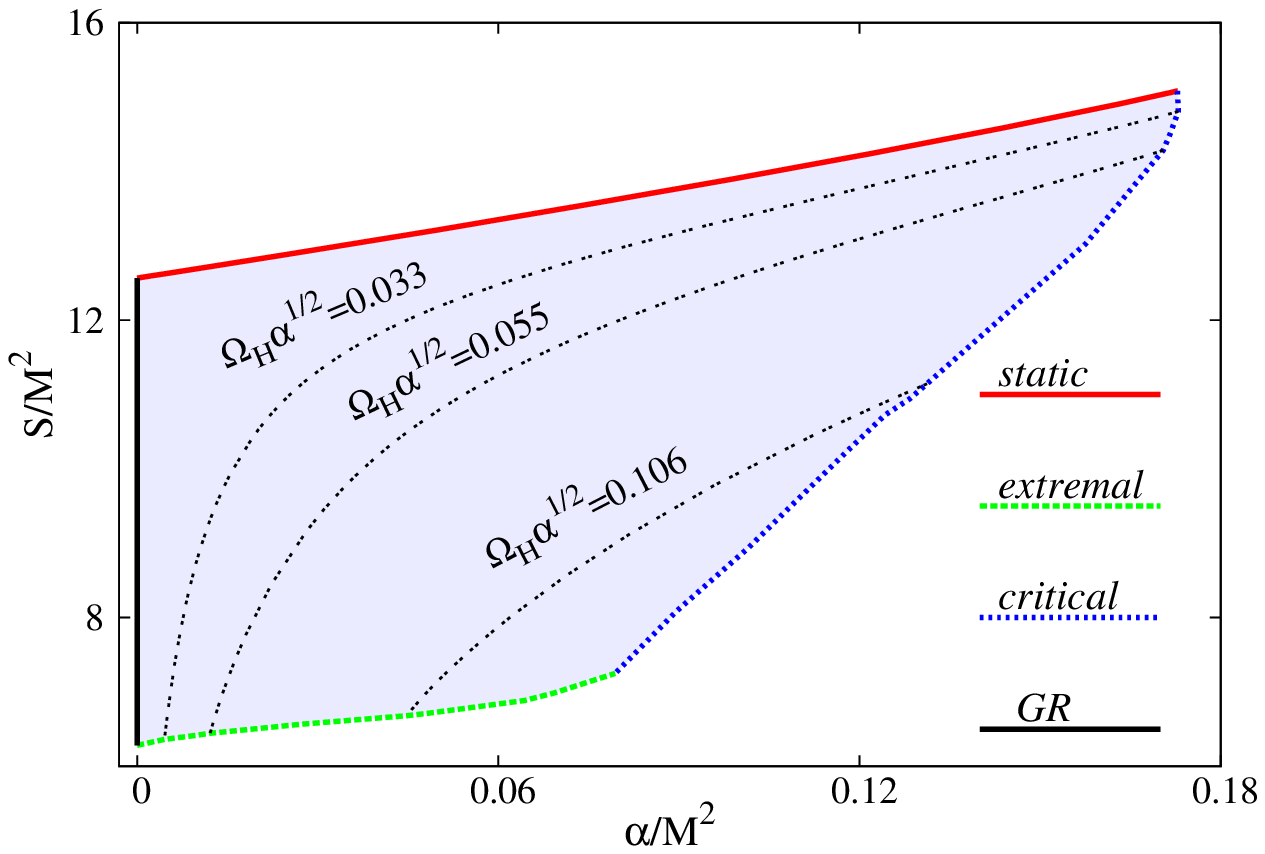} 
\includegraphics[height=.225\textheight, angle =0]{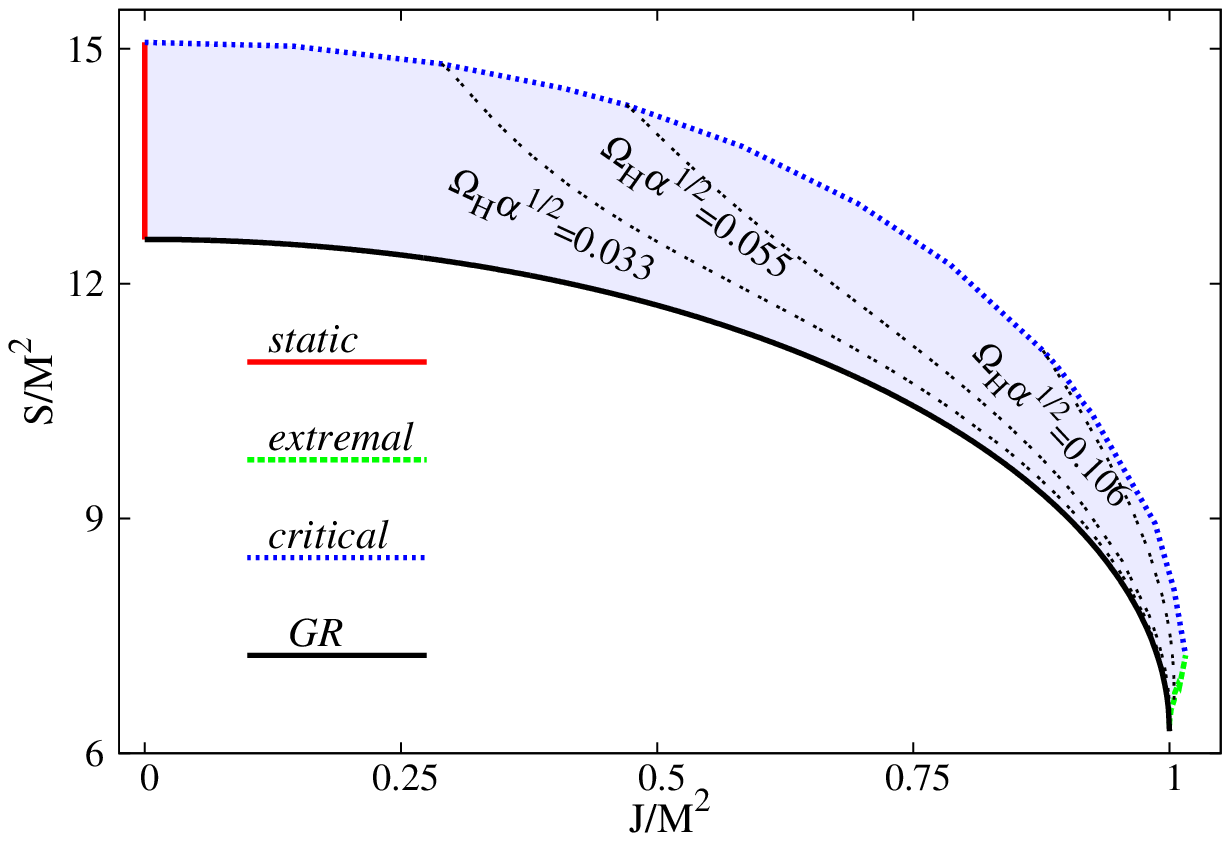} \ \
\includegraphics[height=.225\textheight, angle =0]{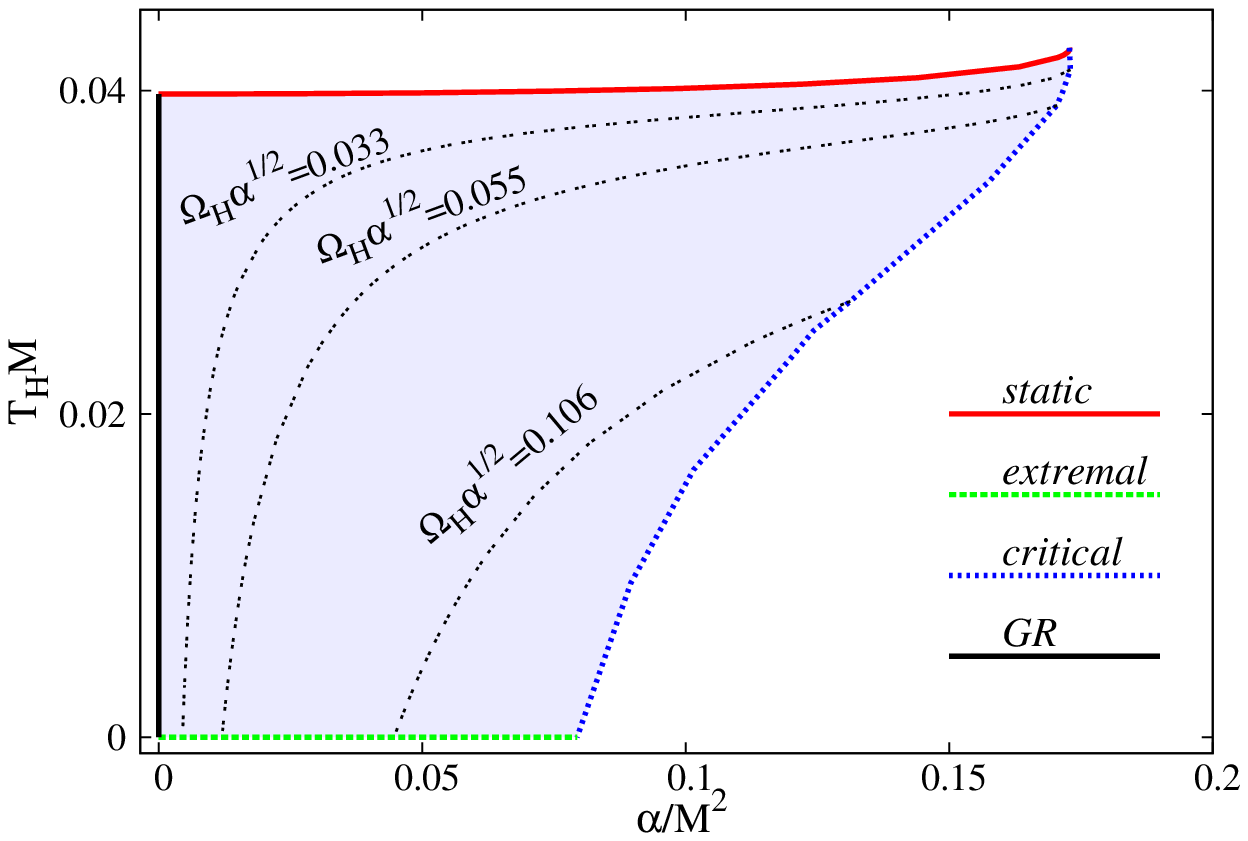} 
\includegraphics[height=.225\textheight, angle =0]{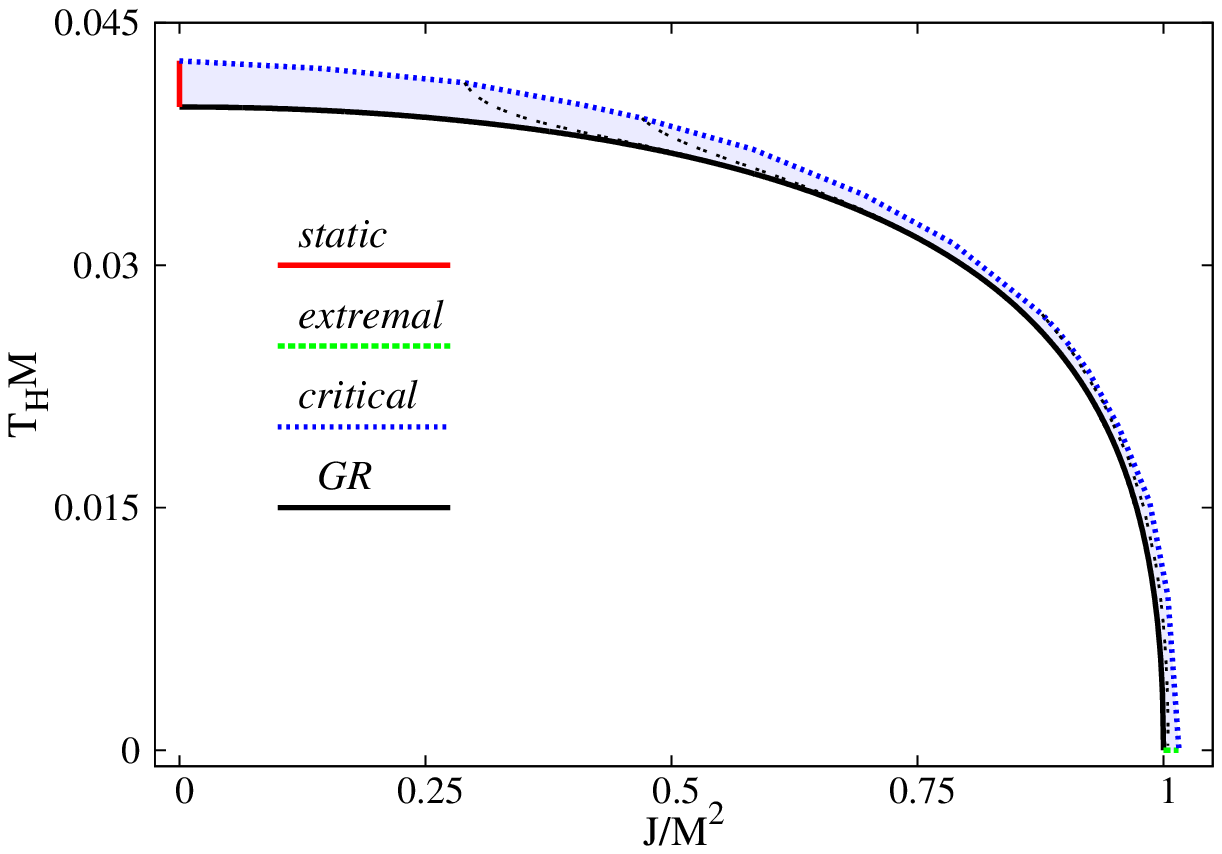} \ \  
\includegraphics[height=.225\textheight, angle =0]{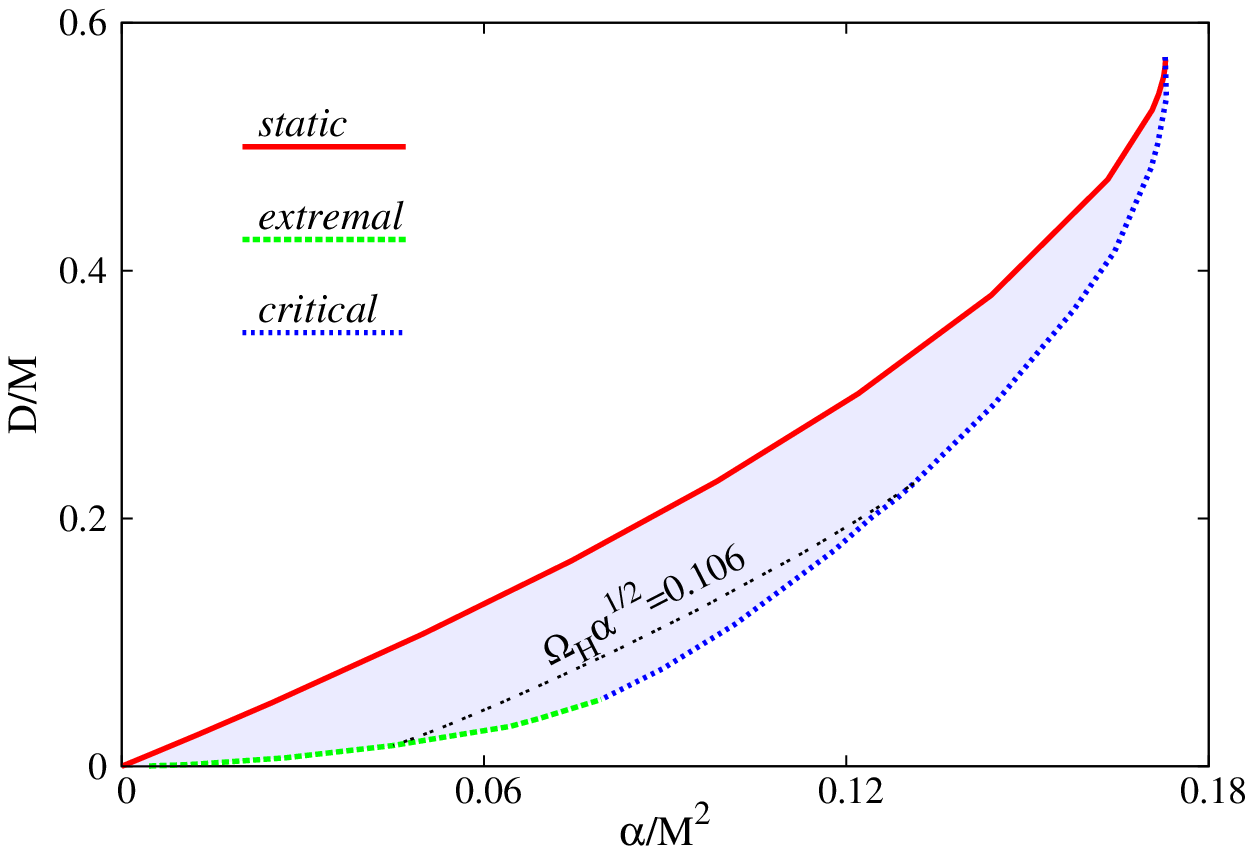} 
\includegraphics[height=.225\textheight, angle =0]{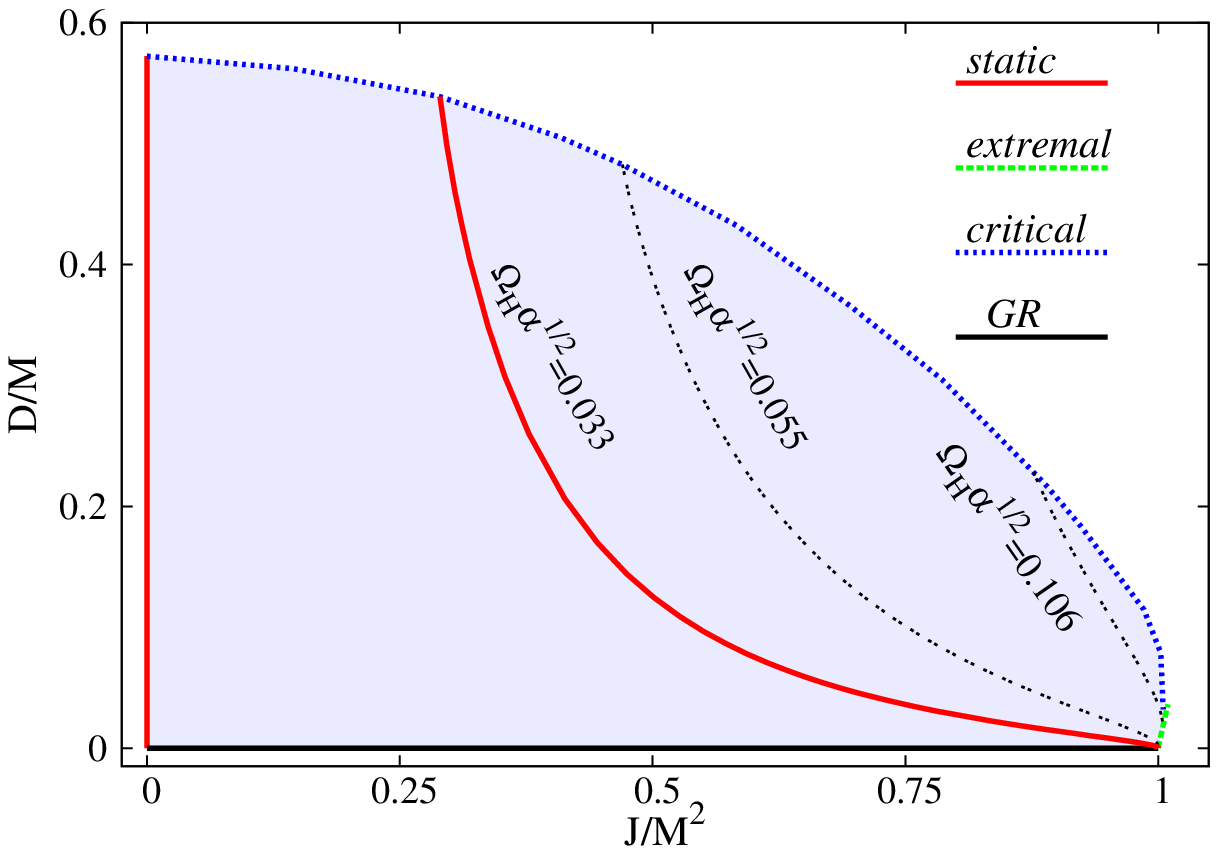} 
\end{center}
  \vspace{-0.5cm}
\caption{  The domain of existence of the horizon area, entropy, temperature and dilaton charge 
is shown vs.~$\alpha$ (left panels) 
and   vs.~$J$ (right panels). Here and in Figure \ref{dom2}, 
all quantities are normalized $w.r.t.$ the mass of the solutions. 
}
\label{dom1}
\end{figure}

\begin{figure}[h!]
\begin{center}  
\includegraphics[height=.28\textheight, angle =0]{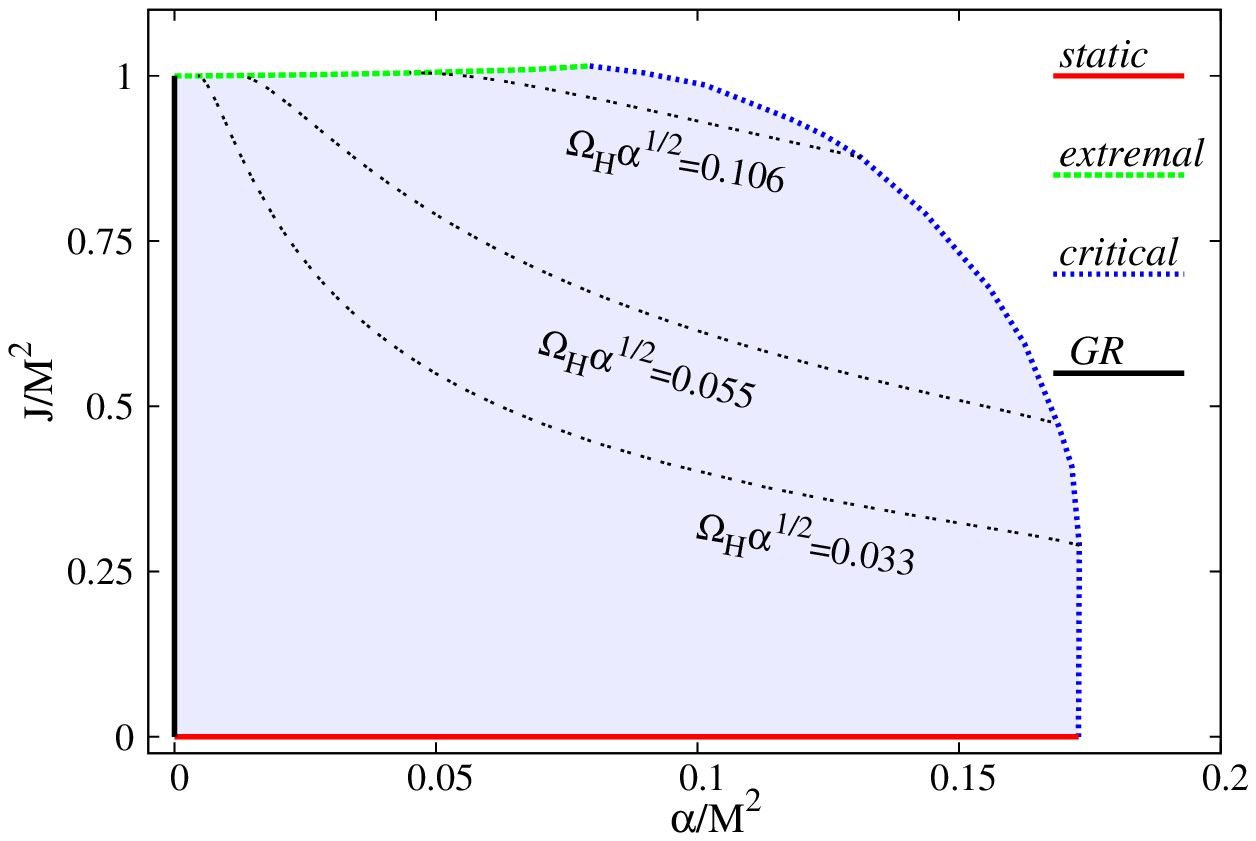} 
\end{center}
  \vspace{-0.5cm}
\caption{ Domain of existence of EGBd  spinning BHs in an angular momentum 
vs.~$\alpha$ diagram.
}
\label{dom2}
\end{figure}
 
\vspace*{0.06cm}

\subsection{The domain of existence }
 
Returning to the discussion of the sets of solutions for
$\Omega_{\rm H}$ and $\alpha$,
let us now address the limiting behavior on the second branch.
As discussed above, the family of Kerr BHs with fixed $\Omega_{\rm H}$ ends in an extremal configuration
with a novanishing horizon area,
which is approached as $r_{\rm H}\to 0$.  
We have found that this applies also for the families of EGBd BHs 
with fixed $\Omega_{\rm H}$ and $\alpha$.
The extrapolated numerical data\footnote{Note, that only near extremal solutions can be constructed within the
framework proposed in this work.} always indicates the existence
of a $r_{\rm H}\to 0$ limiting configuration with vanishing Hawking temperature
and nonvanishing global charges, horizon area and entropy.
However, unlike the extremal Kerr solution,
the extremal EGBd solutions appear to not be regular. 
That is,
the data found for all limiting solutions found so far
show that the dilaton field at the horizon seems to diverge at the poles as $r_{\rm H}\to 0$,
making the extremal solutions singular.
Note, however, that  the (quasi-isotropic) metric functions tend to well defined
limiting functions.

This behaviour is illustrated in Figure \ref{ZH},
where we show how the value of the scalar field at the north pole on the horizon
($\theta=0$) 
changes with the scaled temperature for two values of
the dimensionless parameter $\Omega_{\rm H}\sqrt{\alpha}$.
The curves there start at the corresponding critical solutions 
and end at near-extremal ones (beyond which sufficient numerical accuracy
is lost).
A further argument for the non-existence of regular extremal solutions is
given in Appendix B, and based on an attempt to construct
the near-horizon configurations.

\medskip

Let us now address the domain of existence of the EGBd spinning BHs.
As expected, the general pattern is rather complicated, and depends on the value of
the parameter $\alpha$, 
which sets another length scale of the problem, apart from the mass of the solutions $M$.
 
In Figure \ref{dom1} (left panels) 
the dimensionless horizon area $A_{\rm H}/M^2$, the dimensionless entropy $S/M^2$, 
the dimensionless temperature $T_{\rm H} M$
and the dimensionless dilaton charge $D/M$ of all solutions are
shown
as functions of the dimensionless coupling constant $\alpha/M^2$.
A complementary picture is found when exhibiting the same data as functions of
the dimensionless reduced angular momentum $J/M^2$, 
as shown in Figure \ref{dom1}  (right panels).
Note, that in the Figure all physical quantities are expressed 
in units set by the mass of the solutions.
The link between these two sets is provided by the Figure \ref{dom2}, 
where  we show the dimensionless reduced angular momentum $J/M^2$ 
as a function of the dimensionless coupling constant $\alpha/M^2$.
In these plots we include also the lines corresponding to three constant values of the dimensionless parameter
$\Omega_{\rm H} \sqrt{\alpha}$, which provides a measure of how fast the solutions spin.

Also, let us
 mention that in these plots, the shaded region was obtained by extrapolating to the continuum
the results from a set of around one thousand numerical solutions.

\newpage
In all plots, the region where EGBd spinning BHs exist is delimited by
\begin{description}
\item[i)]   the set of static EGBd BHs (red solid line);
\item[ii)]  the set of extremal ($i.e.$, zero temperature) EGBd BHs  (green dashed line),
which is found by extrapolating the data for near-extremal configurations;
\item[iii)] the set of critical solutions (blue dotted line);
\item[iv)] and the set of GR solutions -- the Kerr BHs (black solid line).
\end{description}

\medskip

These Figures exhibit a number of interesting features of the spinning EGBd solutions:
\begin{itemize}
\item[a)]
For a given value of $\alpha$, the lower bound (\ref{cond}) on the 
BH mass still holds, irrespective of the value of the angular momentum.
Moreover, the solutions saturating this bound have $J=0$.
\item[b)]
For
a given mass and angular momentum,
an EGBd BH has a higher entropy and temperature than a Kerr BH;
however, an EGBd BH has a lower horizon area than a Kerr BH, except
close to the Kerr bound.

\item[c)]
The critical static solution has 
the largest scaled entropy
possible for an EGBd BH, 
while the critical extremal solution has
the smallest scaled area.
The dilaton charge  
is also  maximal for the critical static solution.
\item[d)]
Perhaps the most interesting feature one can notice in Figures 
\ref{dom1} and
\ref{dom2}
is that 
the EGBd spinning BHs can violate the Kerr bound $J/M^2\leq 1$, for a particular 
set of solutions close to extremality.
This violation is rather small\footnote{Much larger violations of the Kerr bound
have been found recently in a model containing a
massive complex scalar field minimally coupled to Einstein gravity
\cite{Herdeiro:2014goa,Herdeiro:2015gia,Kleihaus:2015iea} 
(see also the more general discussion in \cite{Herdeiro:2015moa}). 
There, the violation can be related to the existence of a solitonic limit of the BHs, which is not the case
for the EGBd model.
We conjecture that the violation of the Kerr bound for the solutions in this work
can be attributed to the existence of regions with a negative effective energy density, as obtained from (\ref{Teff}).
(A discussion of this feature
is given in \cite{Kanti:1995vq} for the spherically symmetric limit.)
} (with a maximal value of $J/M^2\simeq 1.03$).
However, this violation clearly exists,
being manifestly above the numerical accuracy of the calculations.
\item[e)]
Somehow unexpected, the maximal value of  
$J/M^2$ is found for a value of $\alpha/M^2 \simeq 0.079$ which is almost in the middle of the allowed interval.
At this special point of the domain of existence,
where the scaled angular momentum reaches its maximal value,
the branches of critical EGBd BHs and singular extremal solutions merge. 
At the same time, the scaled horizon angular velocity reaches there
its maximal value, $\Omega_{\rm H} \alpha^{1/2} \approx 0.135-0.14$.
 
\end{itemize}

\begin{figure}[h!]
\begin{center}
\includegraphics[height=.26\textheight, angle =0]{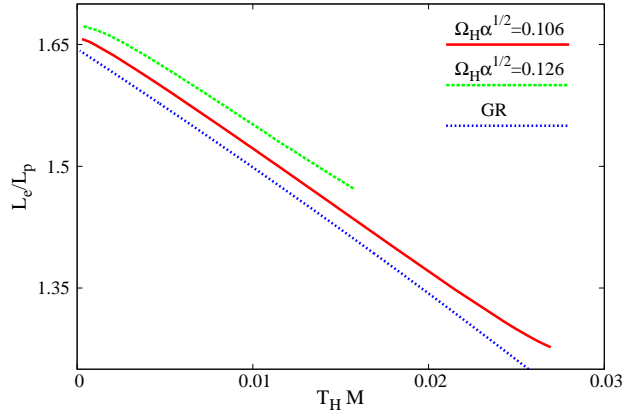} 
\end{center}
  \vspace{-0.5cm}
\caption{  The ratio $L_e/L_p$ is shown for two sets of EGBd solutions with fixed values of
the dimensionless parameter $\Omega_{\rm H}\sqrt{\alpha}$, which
extend from the critical configurations to the extremal ones.
Also shown is this ratio for a set of Kerr solutions
close to extremality.}
\label{LeLp}
\end{figure}
\begin{figure}[h!]
\begin{center}
\includegraphics[height=.26\textheight, angle =0]{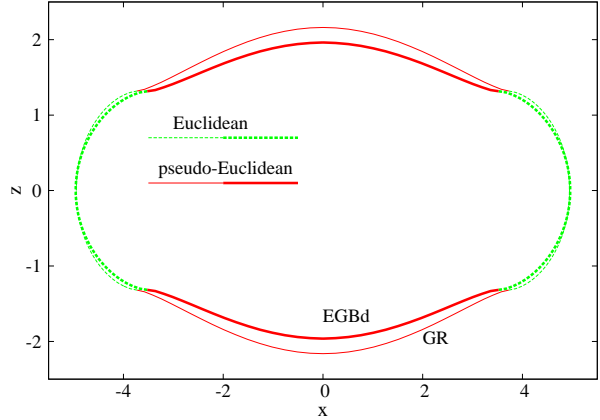} 
\end{center}
  \vspace{-0.5cm}
\caption{  The shape of the horizon is shown for the EGBd 
black hole in Figures \ref{sol1} and \ref{sol2}
and for a Kerr black hole with
the same values of the mass and angular momentum.
The solid lines indicate an embedding in pseudo-Euclidean space, the dashed lines in  Euclidean space.
}
\label{shape}
\end{figure}

\subsection{Further properties }

\subsubsection{Horizon  and ergoregion }
As mentioned above, similar to the Kerr solution in GR, 
these EGBd BHs have an event horizon of spherical topology\footnote{This can be seen, $e.g.$,
from the expression (\ref{horizon-metric}). 
Note that the functions $m_2(\theta)$, $l_2(\theta)$ and $f_2(\theta)$ are strictly positive.
However, one should remark that it is not apriori clear that the all BHs in EGBd theory
would necessarily possess a horizon with $S^2$ topology, 
since a number of classical GR theorems do not apply to this model. }.
Geometrically, however, the
horizon is a squashed sphere. 
This can be seen by evaluating the circumference of the horizon along the
equator, $L_e$, and the circumference of the horizon along the poles, $L_p$

\begin{figure}[h!]
\begin{center}
\includegraphics[height=.26\textheight, angle =0]{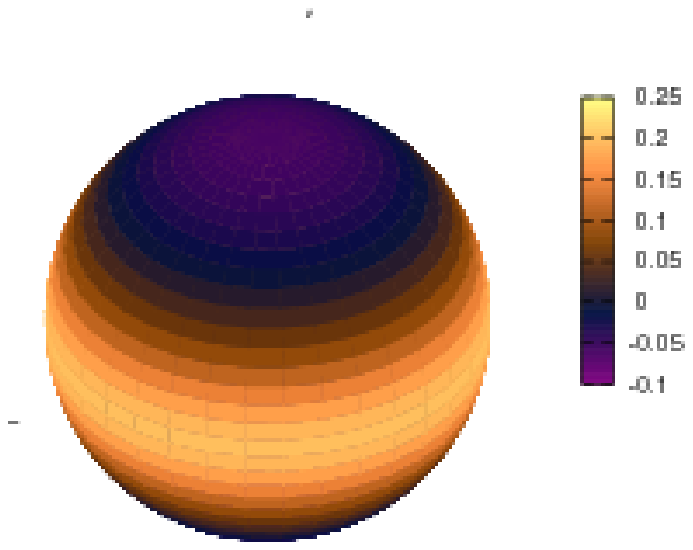} 
\includegraphics[height=.26\textheight, angle =0]{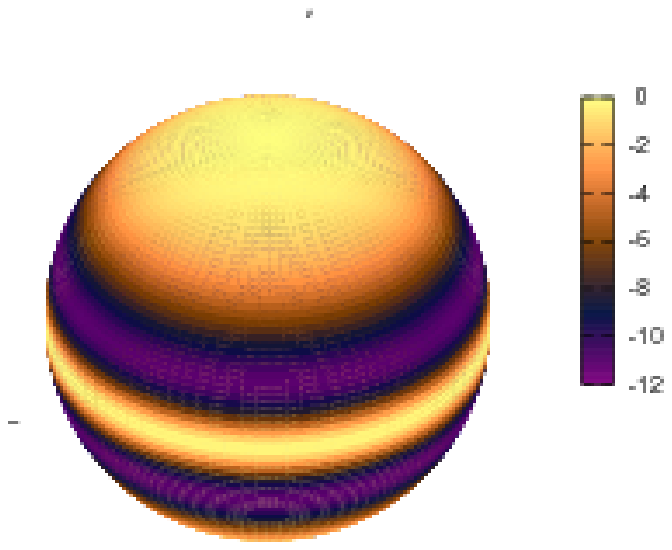} \ \ 
\end{center}
  \vspace{-1.25cm}
\caption{The scalar field (left panel) and its component $T_t^{t(\phi)}$ of the energy momentum tensor 
(multiplied with a factor of $10^3$;
right panel) on the horizon, are shown for the typical EGBd spinning solution 
of Figures \ref{sol1} and \ref{sol2}.}
\label{hor}
\end{figure}
%
\begin{figure}[h!]
\begin{center}
\includegraphics[height=.24\textheight, angle =0]{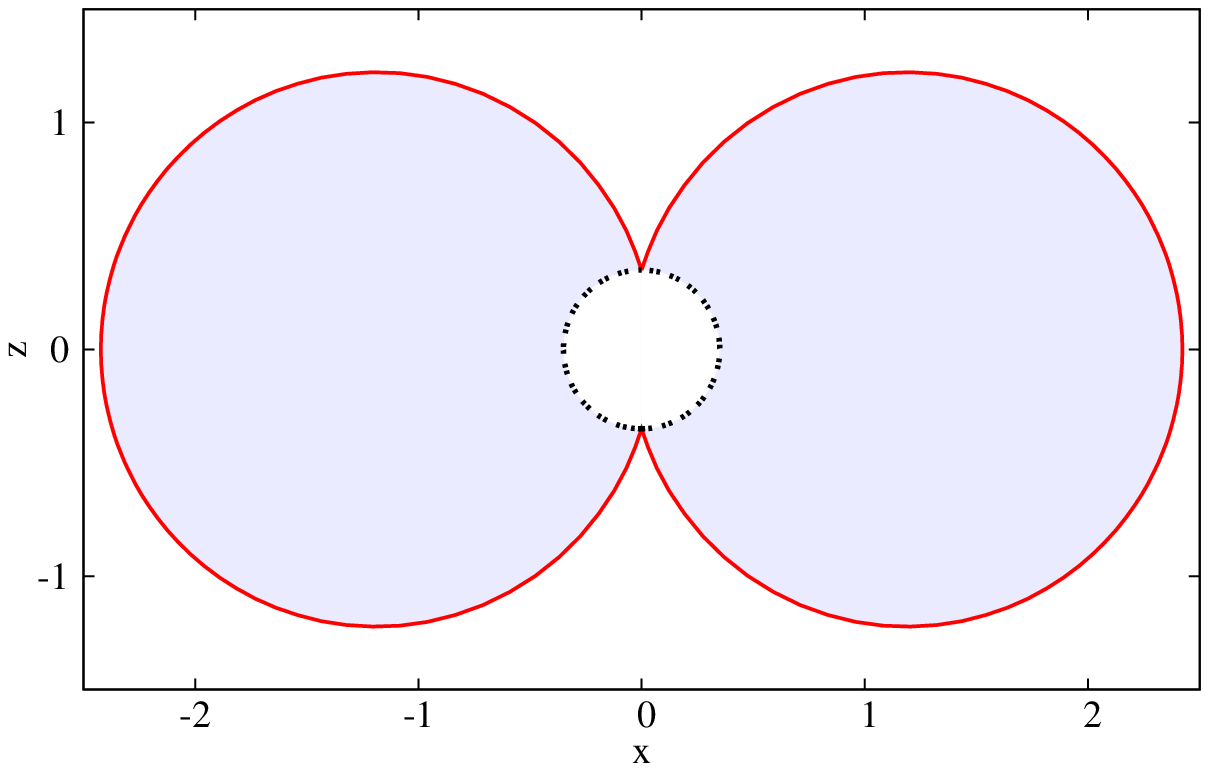} 
\includegraphics[height=.24\textheight, angle =0]{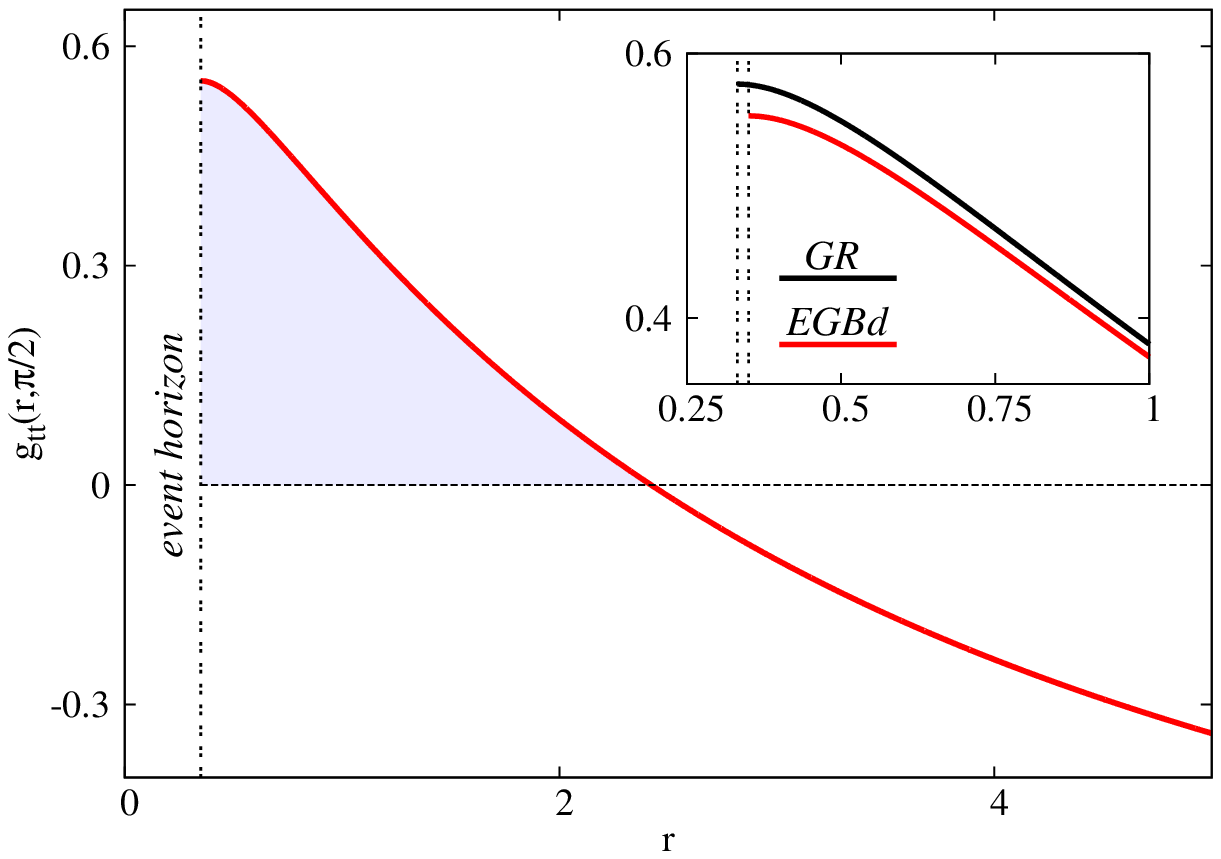} 
\end{center}
\caption{A cross section of the ergosurface  in the $x - z$ plane (left panel) and the metric function $g_{tt}$ in the
equatorial plane (right panel) for the EGBd spinning solution 
in Figures \ref{sol1} and \ref{sol2}. 
In the  right panel, the inset shows the near horizon behaviour of
 $g_{tt}(r,\pi/2)$ for this solution and for a Kerr black hole with
the same values of the mass and angular momentum.}
\label{ergo}
\end{figure}

\begin{eqnarray}
L_e=2\pi r_{\rm H} \sqrt{\frac{l_2(\theta)}{f_2(\theta)}}\bigg|_{\theta=\pi/2},
~~
L_p=2r_{\rm H}\int_0^\pi d\theta \sqrt{\frac{m_2(\theta)}{f_2(\theta)}}.
\end{eqnarray}
In Figure \ref{LeLp} we show the ratio of the equatorial circumference to the polar circumference for several
values of the dimensionless quantity $\Omega_{\rm H}\sqrt{\alpha}$. 
As expected, the squashing of the horizon produced by the rotation is such that $L_e/L_p$ is
always larger than one.

Further insight into the issue of horizon shape is obtained by considering the
 isometric embeddings of
the horizon in a Euclidean space \cite{Smarr:1973zz}.
In Figure \ref{shape} this is done for the solution shown in Figures \ref{sol1} and \ref{sol2}
together with a Kerr BH with
the same values of the mass and angular momentum.
Note that for both GR and EGBd solutions,  the 
embedding of the  horizon cannot be performed completely in Euclidean space with the metric $ds^2 = dx^2 + dy^2 + dz^2$, 
and a region 
(represented by solid lines in that Figure) 
must be embedded in a pseudo-Euclidean space with the metric $ds^2 = dx^2 +dy^2-dz^2$.

On the horizon the scalar field $\phi(r_{\rm H},\theta)$ and the component $T_t^{t(\phi)}$
of the scalar field energy-momentum tensor
 vary with the angular coordinate, as it is manifest in Figure \ref{hor}.
For example, one can see that  $T_t^{t(\phi)}$ vanishes both at the poles and 
in the equatorial plane. 
Also, the scalar field by itself does not contribute to the angular momentum
of these BHs, since
$T_\varphi^{t(\phi)}=0$ everywhere\footnote{
It is interesting to note that this behavior
strongly contrasts with the results found for the spinning BHs
with scalar hair  in 
\cite{Herdeiro:2014goa,Herdeiro:2015gia,Kleihaus:2015iea}.
There, the scalar field is complex and carries a Noether charge, whose density
is proportional to the angular momentum density.
Also, for those solutions  $T_t^{t(\phi)}$ does not vanish in the equatorial plane.}.

We have found that all BHs have an ergoregion, defined as the domain in which $\xi_\mu \xi^\mu$ is positive
(where we recall $\xi=\partial/\partial t$).
This region 
is bounded by the event horizon and by the surface(s) where
\begin{eqnarray}
-f +\sin^2\theta \frac{l}{f} \omega^2 = 0 \ .
 \label{rel-ergo}
\end{eqnarray}

For the Kerr spacetime,  this surface has a spherical topology and
touches the horizon at the poles.
As discussed in  
\cite{Herdeiro:2014jaa,Kleihaus:2015iea},
the ergoregion can be more complicated for BHs with scalar hair, 
with the possible existence of an additional $S^1\times S^1$ ergo-surface (ergo-torus).
However, we have found that this is not the case for EGBd BHs, 
where all solutions are Kerr-like and possess a single topologically $S^2$ 
ergosurface\footnote{This is not a real surprise, since the presence of
an ergo-torus for the solutions in \cite{Herdeiro:2014goa,Kleihaus:2015iea},
can be traced back to the existence of a (spinning) solitonic limit
with an ergo-torus \cite{Kleihaus:2007vk},
which is not the case for
the EGBd theory.}.
In Figure \ref{ergo} (left) we show a cross section of the ergosurface 
for the (typical) solution exhibited in Figures \ref{sol1} and \ref{sol2}.
The metric function $g_{tt}$ in the
equatorial plane is also shown there (right panel).
The event horizon is displayed as
the black dotted curve and the ergosurface  as the solid
red line, while the ergoregion is shaded in blue.
Also, $x$ and $z$ are defined as
standard Cartesian coordinates in terms of $r$ and $\theta$, which enter (\ref{metric}).

Let us mention also that since the functions $f$, $l$ and $m$
are always strictly positive outside the horizon, 
the Lorentzian signature of the metric is preserved there.
Moreover, $f>0$ implies the existence of a Cauchy surface and
thus the absence of closed timelike curves outside the horizon.

\subsubsection{Quadrupole moment and moment of inertia}

On general grounds, one expects the EGBd spinning BHs to have a number of new phenomenological
properties, leading to deviations from the standard Kerr picture in GR.
Let us start by addressing the issue of the quadrupole moment, $Q$,
which is of particular interest 
since, in principle, it can be measured.
Moreover, it can be used to test  the no-hair property \cite{Loeb:2013lfa}.
As discussed in  
\cite{Kleihaus:2014lba}\footnote{The quadrupole moment is extracted 
following Geroch and Hansen \cite{Geroch:1970cd,Hansen:1974zz}
(see also \cite{Hoenselaers:1992bm,Sotiriou:2004ud})},
the quadrupole moment of EGBd BHs is encoded in the free constants which enter the
large-$r$ expansion of the solutions (\ref{inf-expr}), with\footnote{In
the Kerr case the coefficients are $M_2=-2M ( M^2 - 4 r_{\rm H}^2)/3$ and
$C_1=-2r_{\rm H}^2$.}
\begin{eqnarray}
-Q=-M_2+\frac{4}{3}
\left (
\frac{1}{4}+\frac{C_1}{2M^2}+\frac{D^2}{16M^2}
\right)M^3.
\end{eqnarray}
Since the dilaton field enters the expansion for 
the quadrupole moment via a Coulomb-like term in (\ref{inf-expr}),
${D}/{r}$, and since 
there is no explicit contribution from the GB term,
which decays sufficiently fast,
the resulting expression for the quadrupole moment
coincides with the analogous term in Einstein-Maxwell theory 
(after replacing the electro-magnetic charge $Q_e$ by $D$).

\begin{figure}[h!]
\begin{center}
\includegraphics[height=.27\textheight, angle =0]{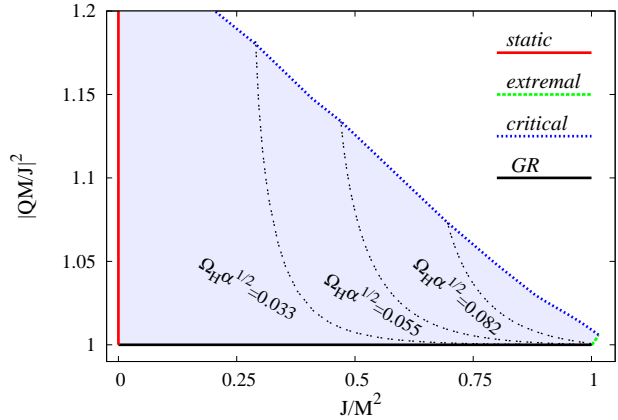} \ \
\end{center}
  \vspace{-0.5cm}
\caption{The magnitude of the scaled quadrupole moment $ QM/J^2$ is shown versus the scaled angular momentum  $J/M^2$ for
several
 values of the scaled horizon angular velocity $\Omega_{\rm H}\sqrt{\alpha}$.
The shaded area represents the domain of existence
of the EBGd black holes.}
\label{quad1}
\end{figure}

In contrast to the EGBd case,
the quadrupole moment of the Kerr solution is completely fixed by 
its global charges, with $Q=-J^2/M$.
In Figure \ref{quad1} the magnitude
of the scaled quadrupole moment $QM/J^2$ is exhibited versus the
scaled angular momentum $J/M^2$.
Here the Kerr solution is represented by the line $|QM/J^2|=1$.
The Figure shows  that the quadrupole moment of
EGBd spinning BHs can take considerably different values from the Kerr
case.
Interestingly, for slow rotation, one finds
deviations from the Kerr value of 20\% and more.
Moreover, the solutions with $J > M^2$, always have
a value of the scaled quadrupole moment greater than one.

\begin{figure}[h!]
\begin{center}
\includegraphics[height=.27\textheight, angle =0]{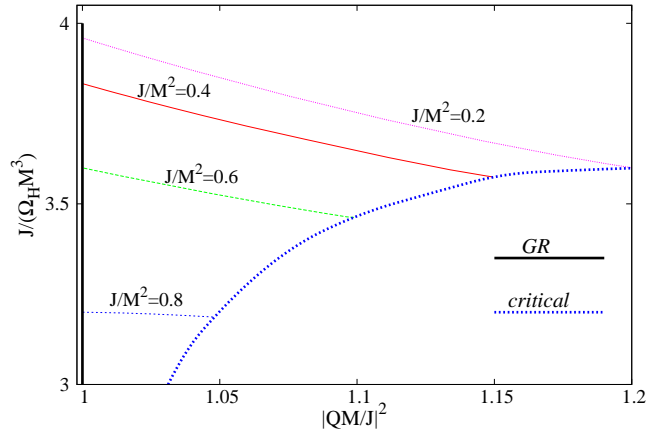} \ \
\end{center}
  \vspace{-0.5cm}
\caption{The scaled moment of inertia $J/(\Omega_{\rm H}M^3)$
is shown versus the magnitude of the scaled quadrupole moment $ QM/J^2$
for several values of the scaled angular momentum  $J/M^2$.}
\label{quad2}
\end{figure}

The scaled moment of inertia 
$J/(\Omega_{\rm H} M^3)$ is exhibited
in Figure \ref{quad2}
versus the scaled quadrupole moment.
The Kerr values are given by
$J/(\Omega_{\rm H} M^3) = 2(1+\sqrt{1-j^2})$ with $j=J/M^2$,
and form the vertical line at $|QM/J^2|=1$,
ranging from the value $J/(\Omega_{\rm H} M^3)=2$ in the extremal
rotating case to the value $J/(\Omega_{\rm H} M^3)=4$, when the
static limit $J\to 0$ is taken.

For the EGBd BHs we have obtained families of solutions
with fixed values of the reduced angular momentum $j=J/M^2$.
For $j\le 1$ these families of solutions start
at their respective Kerr values on the vertical line $|QM/J^2|=1$,
from where they extend, until the respective critical solution is reached.
Thus the Kerr values and the critical values again form two of the
boundaries of the domain of existence.
%
%
%
%
We have not yet mapped out the full domain
of existence, since the limit $j \to 0$ is difficult to obtain from a
numerical point of view, because ratios of small numbers 
are involved.
Here perturbation theory should be helpful, at least
in part of the domain.
Comparing our results with the perturbative results derived
in \cite{Ayzenberg:2014aka} yields good agreement
for small $\alpha$ and small $j$ \cite{Kleihaus:2014lba}.

\subsubsection{Petrov type and ISCOs}

One interesting property of the Kerr metric is that the geodesic equations are separable
due to the existence of the Carter constant \cite{Carter:1968rr}.
This is related to the fact that the Kerr spacetime is  Petrov type D. 
We have investigated the issue of Petrov type for a number of EGBd solutions
in different regions of the parameter space and found that all of them are Petrov type I.
Thus we expect this to be a generic property
of all configurations\footnote{This agrees with the perturbative results in \cite{Ayzenberg:2014aka}.}.
Therefore, the existence of a Carter-like constant is highly unlikely in this case.

\begin{figure}[h!]
\begin{center}
\includegraphics[height=.27\textheight, angle =0]{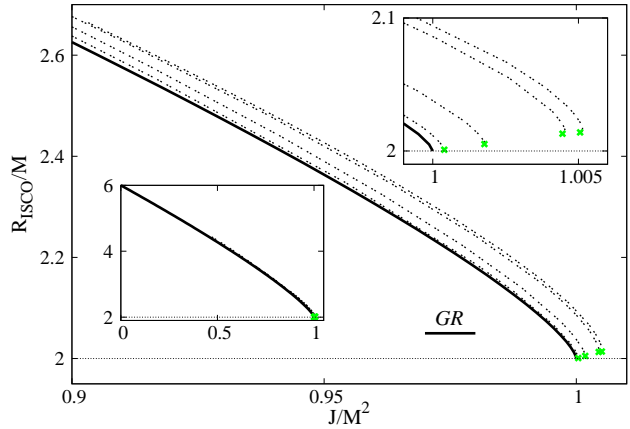} \ \ 
\end{center}
  \vspace{-0.5cm}
\caption{The circumferential ISCO radius $R_{\rm ISCO}$ (given in units of the mass)
is shown
versus the reduced angular momentum $J/M^2$ for several values of the scaled event horizon velocity
$\Omega_{\rm H}\sqrt{\alpha}=0.055$ (the curve closest to the GR one), 
$ 0.082,$ $0.106$  and $0.11$.}
\label{ISCO1}
\end{figure}
 
\begin{figure}[h!]
\begin{center}
\includegraphics[height=.26\textheight, angle =0]{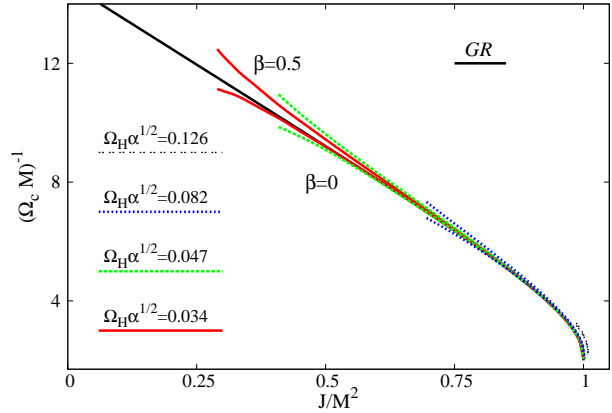}
\end{center}
  \vspace{-0.5cm}
\caption{The inverse of the orbital frequency $\Omega_c$
is shown
versus the angular momentum for coupling constant $\beta=0$ and $\beta=0.5$ 
for several values of the scaled event horizon velocity
$\Omega_{\rm H} \sqrt{\alpha}$.
Both $\Omega_c$ and $J$ are given in units set by the mass.}
\label{ISCO2}
\end{figure}

Clearly, the geodesics of these EGBd BHs 
are also expected to deviate considerably from those of the Kerr BHs.
The geodesics of static and slowly rotating BHs
were considered in \cite{Pani:2009wy}.
There the dependence of the 
innermost-stable-circular-orbits (ISCOs) 
on the angular momentum was studied, showing
that the ISCO is larger for slowly rotating EGBd BHs
than for Kerr BHs, and that 
the orbital frequency is smaller \cite{Pani:2009wy}.

While a systematic discussion of the geodesics
of the general family of rotating EGBd BHs
will be presented elsewhere, 
we here focus on the special case of ISCOs 
for the rapidly spinning BHs \cite{Kleihaus:2011tg}.
The geodesics for circular motion are obtained from
the Lagrangian 
\begin{equation}
\label{ISCO}
2 {\cal L} = e^{-2 \beta \phi} g_{\mu\nu}\dot x^\mu  \dot x^\nu =-\epsilon,
\end{equation}
where $\epsilon=0$ and 1 massless and massive particles, respectively.
The constant $\beta$ fixes the coupling between the matter and the dilaton field.
(In heterotic string theory $\beta=0.5$.)
Also, a superposed dot denotes the derivative with respect to the affine parameter along the geodesics. 
As usual, the existence of the Killing vectors  $\xi$ and $\eta$
implies the existence of two conserved quantities, 
the energy $E$ and the angular momentum $L$ of the particle, respectively.
Restricting to (timelike) motion in the
equatorial plane ($\theta=\pi/2$), one finds from (\ref{ISCO})
the equation
\begin{equation}
\label{ISCO21}
\dot r^2=e^{2\beta \phi }\frac{f}{m}
\left(
e^{2\beta \phi }\frac{\left(E -\frac{L \omega}{r}\right)^2}{f}
-e^{2\beta \phi }\frac{L^2 f}{r^2 l}
-1
\right)  
\equiv V(r).
\end{equation}
Also, the orbital angular velocity is expressed as
 \begin{equation}
\label{ISCO12}
\Omega_c=\frac{\dot \varphi}{\dot t}=\frac{\omega}{r}+\frac{1}{r^2}\frac{Lf^2}{l(E-\frac {L\omega}{r})}.
\end{equation}
 For circular orbits $V(r)=V'(r)=0$.
This results in two algebraic relations for $E$ and $L$
which are solved analytically.
Similar to the Kerr case, 
this gives two distinct pairs of solutions 
corresponding to co-rotating and counter-rotating trajectories.
Stability of the orbits further requires that the second derivative
of the effective potential is negative,
$V''(r) < 0$. This selects a particular value of
the radial coordinate $r=r_{\rm ISCO}$,
which separates the stable circular orbits, $r>r_{\rm ISCO}$,
from the unstable ones. 

The results for EGBd ISCOs for rapidly rotating BHs are
exhibited in Figure \ref{ISCO1} for $\beta=0.5$
together with the corresponding Kerr ISCOs. 
Note, that we exhibit the
circumferential radius $R_{\rm ISCO}=\left. r\sqrt{m/f}\right|_{r_{\rm ISCO}}$ here 
(and not the Boyer-Lindquist coordinate radius).
As seen in the Figure, the ISCOs of the EBGd BHs remain larger than the
Kerr ISCOs (as long as the corresponding Kerr BHs exist),
also for rapidly rotating BHs
\cite{Kleihaus:2011tg}. 
Finally, we exhibit in Figure \ref{ISCO2}
the inverse of the orbital frequencies $\Omega_c$ for the EGBd ISCOs 
for $\beta=0.5$ and for $\beta=0$ and compare again with the Kerr case.
For $\beta=0.5$ the frequencies are smaller than the Kerr frequencies.
Here rather large deviations can occur, amounting up to 60\%  close 
to $J/M^2=1$\cite{Kleihaus:2011tg}.
For $\beta=0$ on the other hand the deviations are smaller and exceed the Kerr values.

\section{ Conclusions }

 The main purpose of this work was to provide a more detailed description of the construction and of the  
(basic)
physical properties of the spinning EGBd BHs reported in \cite{Kleihaus:2011tg}.
These configurations can be viewed as the counterparts of Kerr solutions in the  
presence of a dilatonic GB term in the gravity action. Our results here, together with those in 
\cite{Kleihaus:2011tg,Kleihaus:2014lba}
show that such BHs share the basic properties of the GR BHs.
However, new features occur as well.
The most interesting results of this work are perhaps the ones exhibited in the
Figures \ref{dom1} and \ref{dom2},
which exhibit the domain of existence of the EGBd solutions,
both in terms of the GB coupling constant and in terms of the angular momentum.
In particular, we find that similar to the static case,
the spinning EGBd  BHs possess a minimal value of the mass.
Moreover, their angular momentum can slightly exceed the Kerr bound.

\medskip
This work can be continued in many possible directions.
Perhaps the most
important one is to clarify the issue of stability.
For spherically symmetric solutions, (mode) stability was established in \cite{Kanti:1997br} for radial
perturbations and in \cite{Pani:2009wy} for axial perturbations.
However, due to the complicated form of the field equations, extending this study to
the spinning case will be a highly nontrivial task.
In this context, let us remark that, as discussed above, these spinning BHs have an ergoregion.
Thus, similar to the Kerr case, they should be afflicted by superradiant instabilities 
in the presence of (massive)
bosonic fields.

Another interesting direction 
would be to further explore astrophysical signatures of these EGBd BHs.
An obvious task here will be to study the geodesics in a systematic way
and to compute the shadows of EGBd BHs,
contrasting the results with
those for the Kerr solution\footnote{For example, the results in the recent work 
\cite{Cunha:2015yba}
show that the shadow of a spinning BH with scalar hair (for a model 
with a massive complex scalar field)
can be drastically different
as compared to the GR results.
}.

On the more theoretical side,
it would be interesting to  extend the solutions in this work
to other values of the coupling constant $\gamma$, which enters the exponent in (\ref{act}),
 and to look for generic properties.
Moreover, 
one  may also search for new BH solutions in more general theories containing (\ref{act}) as the basic piece.
The most obvious example here would be to consider generalizations of
the Kerr-Newman BHs in EGBd-Maxwell theory.
Also, it would be interesting to consider rotating solutions with different asymptotics,
$e.g.$ (anti)-de Sitter asymptotics, by supplementing (\ref{act}) with a cosmological term.

 \vspace*{0.5cm}
\noindent{\textbf{~~~Acknowledgements.--~}} 
We acknowledge  support by the DFG Research
Training Group 1620 ``Models of Gravity”.
B.K.~and J.K.~gratefully acknowledge support from FP7,
Marie Curie Actions, People, International Research Staff Exchange Scheme 
(IRSES-606096).
  E.R. gratefully acknowledges support from the FCT-IF programme.

\appendix

\section{The critical solutions}

As noted in Section 3.1,
  the spinning solutions BH on the fundamental branch 
cease to exist for $r_{\rm H}$ smaller
than some critical value $r_{\rm H}^{(cr)}(\alpha,\Omega_{\rm H})$. 

\begin{figure}[h!]
\begin{center}
\includegraphics[height=.26\textheight, angle =0]{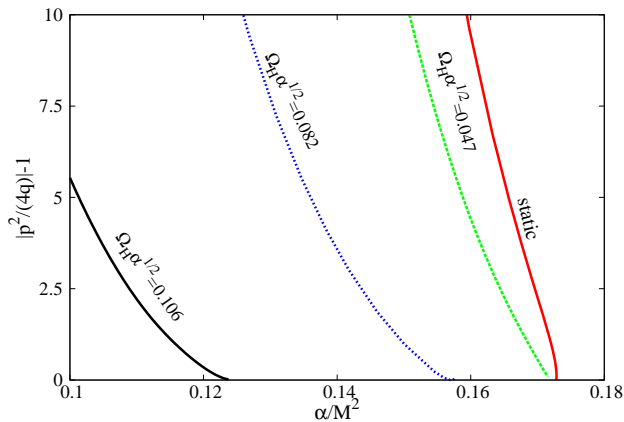} 
\end{center}
  \vspace{-0.5cm}
\caption{  The quantity $|p^2/(4q)|-1$ which enters the discriminant $\Delta$
of the equation (\ref{eq1})
is shown as a function of $\alpha/M^2$ for several values of the scaled angular
velocity $\Omega_{\rm H}\sqrt{\alpha}$.
}
\label{end}
\end{figure}

To understand this behaviour, one has to return to 
 the horizon expansion of the metric functions (\ref{horizon-expansion}), 
and look for the expressions of higher order terms.
Then, after some algebra, one finds that the second order term $\phi_2(\theta)$ in the expansion of the 
scalar field $\phi(r,\theta)=\phi_0(\theta)+\phi_2(\theta) (r/r_{\rm H}-1)^2+\dots$
is, similar to the static limit, a solution of a quadratic equation,
\begin{eqnarray}
\label{eq1}
 \phi_2^2+p \phi_2+q=0,
\end{eqnarray}
where the coefficients $p$ and $q$ depend on $f_2(\theta)$, $l_2(\theta)$, 
and $m_2(\theta)$,
and on their first and second derivatives\footnote{We recall that the ratio 
$f_2^2/m_2$ is constant (see relation (\ref{TH})).}.
Then a real solution to the above equation exists only if $\Delta=p^2-4 q>0$. 
In practice, we have monitored this discriminant for a set of values of $(\alpha,\Omega_{\rm H})$ and
observed that the solutions always cease to exist exactly when $\Delta$ becomes 
negative. This is illustrated in Figure \ref{end}.\footnote{We
note, that for the spinning solutions we did not observe 
a minuscule branch whose mass increases when the horizon
size decreases, as present in the static case.
While by continuity, such a branch should be present for
sufficiently slowly rotating BHs, its uncovering would need
rather high numerical accuracy.}

We mention that the analogous behaviour has been noticed for other non-spherically symmetric solutions
with a GB term in the action (see \cite{Kleihaus:2009dm,Kleihaus:2012qz}).

\section{The issue of extremal solutions}
  
 As discussed in Section 3, the numerical solutions with  
very small values of the Hawking temperature 
have smooth metric functions which tend to well defined
limits as $T_{\rm H}\to 0$; however, 
their scalar field takes large values at the poles
of the BH horizon.
Thus, by extrapolating this result, it is natural
to conjecture
that the extremal solution is singular, with a divergent scalar field.
 
A partial confirmation of this conjecture, together with
some analytical understanding, can be
achieved 
when instead of solving the full bulk EGBd equations searching for extremal solutions,
 one attempts to 
construct the corresponding 
near-horizon configurations.
 There one deals with a codimension one problem, whose solutions are easier to study.

In this approach, following $e.g.$ \cite{Astefanesei:2006dd},
one considers the following line element
 with an isometry group $SO(2,1)\times U(1)$:
\begin{eqnarray}
\label{metric-nh}
 ds^2=v_1(\theta)^2 \left ( -r^2 dt^2+\frac{dr^2}{r^2}+\beta^2 d\theta^2 \right)
 +\beta^2 v_2^2(\theta) \left(d\varphi- K r dt \right)^2 ,
\end{eqnarray}
 where 
$\beta,~K$
 are real parameters,
$0\leq r<\infty$, while $\theta,\varphi$ and $t$ have the usual range.
The above line element would
 describe the
neighbourhood of the event horizon of an extremal EGBd BH\footnote{It is interesting to note that, as discussed in 
\cite{Blazquez-Salcedo:2013muz}
in a different context,
the existence of the near horizon solution does not guarantee 
 the existence of a corresponding bulk configuration.
 } (and will 
be an attractor for the full bulk solutions).
 
  The  equations satisfied by $v_i$ can be derived from the reduced Lagrangian
 \begin{eqnarray}
 \label{Lg}
&&L=-2 \beta^2v_1 v_2+\beta^4 \frac{K^2 v_2^3}{2 v_1}
+\frac{2v_2 v_1'^2}{v_1}+4 v_1'v_2'
-\frac{1}{2}v_1v_2 \phi'^2
\\
\nonumber
&&~~~+2\alpha\gamma e^{-\gamma \phi}\frac{\phi'}{v_1}
\bigg(
2\beta^2 K^2 \frac{v_2^3v_1'}{v_1^3}+4v_2'
-3\beta^2K^2 \frac{v_2^2v_2'}{v_1^2}+\frac{4v_1'^2v_2}{\beta^2 v_1^2}
\bigg),
\end{eqnarray}
where a prime denotes a derivative $w.r.t.$ the angular variable $\theta$.
Without any loss of generality,  we set $K=1$ in the above relations\footnote{This is a result of the scaling symmetry
$v_2\to \lambda v_2,~~K\to K/\lambda$.}.

The Einstein gravity solution is found for $\alpha=0$, $\phi=const.$ and reads  \cite{Horowitz}
 \begin{eqnarray}
\label{solE}
 \beta=1,~~v_1^2=\frac{J}{32 \pi}(3+2\cos 2\theta),~~v_2^2=\frac{J}{4 \pi}\frac{ \sin^2\theta}{(1+\cos^2 \theta)},
 \end{eqnarray}
 with $J$ an arbitrary parameter.
 
Although we have failed to find an exact solution in the general  EGBd case, 
a solution can be constructed perturbatively in $\alpha$ around the configuration (\ref{solE}),  by taking
 \begin{eqnarray}
\label{sol-pert}
 \beta=1+\alpha \beta_1+ \dots,~~
v_1(\theta)=v_{10}(\theta)+\alpha v_{11}(\theta) +\dots,
\\
\nonumber
v_2(\theta)=v_{20}(\theta)+\alpha v_{21}(\theta) +\dots,~~
\phi(\theta)= \phi_0+\alpha \phi_{1}(\theta) +\dots,
 \end{eqnarray}
 where $v_{10}(\theta)$ and $v_{20}(\theta)$ are the Einstein gravity functions.
By solving a linear ordinary differential equation, one finds the following expression for
 $\phi_{1}(\theta)$:
  \begin{eqnarray}
\label{phi1}
\phi_{1}(\theta)=
 c_1\log (\tan \frac{\theta}{2})+c_2+64 e^{-\gamma \phi_0} \gamma \pi \frac{(11+20 \cos 2\theta +\cos 4\theta)}{J(3+\cos 2\theta)^2}
\\
\nonumber
-32 e^{-\gamma \phi_0} \frac{\gamma \pi}{J}\log (3+\cos 2 \theta)+64 e^{-\gamma \phi_0}  \frac{\gamma \pi}{J}\log (\sin \theta),
 \end{eqnarray}
with $c_1$, $c_2$ constants of integration.
 One can easily see that, for any choice of the constants $c_1$, $c_2$ the function $\phi_{1}(\theta)$
 necessarily diverges at $\theta=0$ and/or $\theta=\pi$.

 One may argue that this is an artifact of the first order perturbation theory.
 However, we have failed to  find nonperturbative (numerical) solutions with the required asymptotics.
 Let us briefly describe our approach.
The approximate form of the (putative) solutions as $\theta\to 0$ is
   \begin{eqnarray}
\label{t0}
 &&v_1(\theta)=v_{10}-\frac{1}{4}\beta^2 v_{10}\theta^2+\dots,~~
 v_2(\theta)=v_{10}\theta +\frac{1}{12}\beta^2 v_{10}\theta^3+\dots,~~
 \\
 \nonumber
&&\phi(\theta)=\phi_0+\frac{9\alpha \gamma \beta^4e^{\gamma\phi_0}v_{10}^2}{192
 \alpha^2\gamma^2-4e^{2\gamma\phi_0}v_{10}^4 }\theta^4+\dots.
 \end{eqnarray}
A similar expression holds as $\theta \to \pi$
(here we do not suppose the existence of a reflection symmetry in the equatorial plane),
with
\begin{eqnarray}
\label{tpi}
 &&v_1(\theta)=\bar v_{10}-\frac{1}{4}\beta^2 \bar v_{10}(\pi-\theta)^2+\dots,~~
 v_2(\theta)=\bar v_{10}(\pi-\theta) +\frac{1}{12}\beta^2 \bar v_{10}(\pi-\theta)^3+\dots,~~
 \\
 \nonumber
&&\phi(\theta)=\bar \phi_0+\frac{9\alpha \gamma \beta^4e^{\gamma\bar \phi_0}\bar v_{10}^2}{192
 \alpha^2\gamma^2-4e^{2\gamma\bar \phi_0}\bar v_{10}^4 }(\pi-\theta)^4,
 \end{eqnarray}
 where $v_{10}$ and $\bar v_{10}$ are constants.
 In this approach, the shooting parameter is $\beta$ and the solutions
 should exist for a finite range of $v_{10}$ and/or $\bar v_{10}$.
 
In our attempt, one starts with the asymptotics (\ref{t0}) for some fixed $v_{10}$
and tries to adjust the parameter $\beta$ such that (\ref{tpi})
is smoothly approached as $\theta\to  \pi$.
 However,  for all configurations we have found so far, this was not the case.
 The scalar field $\phi(\theta)$ is a monotonic function,
 having a tendency to diverge as $\theta \to \pi$.
This can be understood from the small $\alpha$ expansion (\ref{phi1}),
 which thus remains valid also within a non-perturbative approach.
 
 The natural interpretation of these results is the absence of the EGBd  solutions with a line element 
of the form
(\ref{metric-nh}) and a regular scalar field.
Thus the inclusion of a GBd term in the action 
 does not allow for the rotating  BHs to approach extremality
and remain regular.



\begin{thebibliography}{99}
\bibitem{Moura:2006pz}
  F.~Moura and R.~Schiappa,
  Class.\ Quant.\ Grav.\  {\bf 24} (2007) 361
  [hep-th/0605001].
\bibitem{Kobayashi:2011nu}
  T.~Kobayashi, M.~Yamaguchi and J.~Yokoyama,
  Prog.\ Theor.\ Phys.\  {\bf 126} (2011) 511
  [arXiv:1105.5723 [hep-th]].
\bibitem{Berti:2015itd}
  E.~Berti {\it et al.},
  arXiv:1501.07274 [gr-qc].
\bibitem{Kanti:2011jz}
  P.~Kanti, B.~Kleihaus and J.~Kunz,
  Phys.\ Rev.\ Lett.\  {\bf 107} (2011) 271101
  [arXiv:1108.3003 [gr-qc]].
\bibitem{Kanti:2011yv}
  P.~Kanti, B.~Kleihaus and J.~Kunz,
  Phys.\ Rev.\ D {\bf 85} (2012) 044007
  [arXiv:1111.4049 [hep-th]].
	
\bibitem{Mignemi:1992nt}
  S.~Mignemi and N.~R.~Stewart,
  Phys.\ Rev.\  D {\bf 47} (1993) 5259
  [arXiv:hep-th/9212146].
\bibitem{Mignemi:1993ce}
  S.~Mignemi,
  Phys.\ Rev.\  D {\bf 51} (1995) 934
  [arXiv:hep-th/9303102].
\bibitem{Kanti:1995vq}
  P.~Kanti, N.~E.~Mavromatos, J.~Rizos, K.~Tamvakis and E.~Winstanley,
  Phys.\ Rev.\  D {\bf 54} (1996) 5049
  [arXiv:hep-th/9511071].
\bibitem{Torii:1996yi}
  T.~Torii, H.~Yajima and K.~i.~Maeda,
  Phys.\ Rev.\  D {\bf 55} (1997) 739
  [arXiv:gr-qc/9606034].
\bibitem{Alexeev:1996vs}
  S.~O.~Alexeev and M.~V.~Pomazanov,
  Phys.\ Rev.\  D {\bf 55} (1997) 2110
  [arXiv:hep-th/9605106].
\bibitem{Guo:2008hf}
  Z.~K.~Guo, N.~Ohta and T.~Torii,
  Prog.\ Theor.\ Phys.\  {\bf 120} (2008) 581
  [arXiv:0806.2481 [gr-qc]].
\bibitem{Pani:2009wy}
  P.~Pani and V.~Cardoso,
  Phys.\ Rev.\  D {\bf 79} (2009) 084031
  [arXiv:0902.1569 [gr-qc]].
\bibitem{Maselli:2015tta}
  A.~Maselli, P.~Pani, L.~Gualtieri and V.~Ferrari,
  Phys.\ Rev.\ D {\bf 92} (2015) 8,  083014
  [arXiv:1507.00680 [gr-qc]].
\bibitem{Pani:2011gy}
  P.~Pani, C.~F.~B.~Macedo, L.~C.~B.~Crispino and V.~Cardoso,
  Phys.\ Rev.\ D {\bf 84} (2011) 087501
  [arXiv:1109.3996 [gr-qc]].
		
\bibitem{Ayzenberg:2014aka}
  D.~Ayzenberg and N.~Yunes,
  Phys.\ Rev.\ D {\bf 90} (2014) 044066
   [Phys.\ Rev.\ D {\bf 91} (2015) 6,  069905]
  [arXiv:1405.2133 [gr-qc]].
\bibitem{Kleihaus:2011tg}
  B.~Kleihaus, J.~Kunz and E.~Radu,
  Phys.\ Rev.\ Lett.\  {\bf 106} (2011) 151104
  [arXiv:1101.2868 [gr-qc]].
 
\bibitem{Kleihaus:2014lba}
  B.~Kleihaus, J.~Kunz and S.~Mojica,
  Phys.\ Rev.\ D {\bf 90} (2014) 6,  061501
  [arXiv:1407.6884 [gr-qc]].
\bibitem{Yagi:2012gp}
  K.~Yagi,
  Phys.\ Rev.\ D {\bf 86} (2012) 081504
  [arXiv:1204.4524 [gr-qc]].
\bibitem{Wald:rg} 
R.~M.~Wald, \emph{\ General Relativity}, Chicago, Chicago
Univ. Press, (1984).
\bibitem{Kleihaus:2000kg}
  B.~Kleihaus and J.~Kunz,
  Phys.\ Rev.\ Lett.\  {\bf 86} (2001) 3704
  [gr-qc/0012081];
	\\
  B.~Kleihaus, J.~Kunz and F.~Navarro-Lerida,
  Phys.\ Rev.\ D {\bf 66} (2002) 104001
  [gr-qc/0207042].
	
 \bibitem{Wald:1993nt}
  R.~M.~Wald,
  Phys.\ Rev.\  D {\bf 48} (1993) 3427
  [arXiv:gr-qc/9307038].
\bibitem{schoen}
 W. Sch\"onauer and R. Wei\ss ,
 J. Comput. Appl. Math. 27, 279 (1989) 279;
 \\
 M. Schauder, R. Wei\ss\ and W. Sch\"onauer,
 {\it The CADSOL Program Package},
 Universit\"at Karlsruhe, Interner Bericht Nr. 46/92 (1992).	

\bibitem{Herdeiro:2015waa}
  C.~A.~R.~Herdeiro and E.~Radu,
  Int.\ J.\ Mod.\ Phys.\ D {\bf 24} (2015) 09,  1542014
  [arXiv:1504.08209 [gr-qc]].
 
\bibitem{Kleihaus:2009dm}
  B.~Kleihaus, J.~Kunz and E.~Radu,
  JHEP {\bf 1002} (2010) 092
  [arXiv:0912.1725 [gr-qc]].
\bibitem{Smarr:1973zz}
  L.~Smarr,
  Phys.\ Rev.\ D {\bf 7} (1973) 289.
\bibitem{Kleihaus:2012qz}
  B.~Kleihaus, J.~Kunz, E.~Radu and B.~Subagyo,
  Phys.\ Lett.\ B {\bf 713} (2012) 110
  [arXiv:1205.1656 [gr-qc]].
\bibitem{Herdeiro:2014goa}
  C.~A.~R.~Herdeiro and E.~Radu,
  Phys.\ Rev.\ Lett.\  {\bf 112} (2014) 221101
  [arXiv:1403.2757 [gr-qc]].
\bibitem{Herdeiro:2015gia}
  C.~Herdeiro and E.~Radu,
  Class.\ Quant.\ Grav.\  {\bf 32} (2015) 14,  144001
  [arXiv:1501.04319 [gr-qc]].
\bibitem{Kleihaus:2015iea}
  B.~Kleihaus, J.~Kunz and S.~Yazadjiev,
  Phys.\ Lett.\ B {\bf 744} (2015) 406
  [arXiv:1503.01672 [gr-qc]].
\bibitem{Herdeiro:2015moa}
  C.~A.~R.~Herdeiro and E.~Radu,
  arXiv:1505.04189 [gr-qc].
\bibitem{Herdeiro:2014jaa}
  C.~Herdeiro and E.~Radu,
  Phys.\ Rev.\ D {\bf 89} (2014) 12,  124018
  [arXiv:1406.1225 [gr-qc]].
\bibitem{Kleihaus:2007vk}
  B.~Kleihaus, J.~Kunz, M.~List and I.~Schaffer,
  Phys.\ Rev.\ D {\bf 77} (2008) 064025
  [arXiv:0712.3742 [gr-qc]].
\bibitem{Loeb:2013lfa}
  A.~E.~Broderick {\it et al.},
  Astrophys.\ J.\  {\bf 784} (2014) 7
  [arXiv:1311.5564 [astro-ph.HE]].
\bibitem{Geroch:1970cd}
  R.~P.~Geroch,
  J.\ Math.\ Phys.\  {\bf 11} (1970) 2580.
\bibitem{Hansen:1974zz}
  R.~O.~Hansen,
  J.\ Math.\ Phys.\  {\bf 15} (1974) 46.
\bibitem{Hoenselaers:1992bm}
  C.~Hoenselaers and Z.~Perjes,
  Class.\ Quant.\ Grav.\  {\bf 7} (1990) 1819.
\bibitem{Sotiriou:2004ud}
  T.~P.~Sotiriou and T.~A.~Apostolatos,
  Class.\ Quant.\ Grav.\  {\bf 21} (2004) 5727
  [gr-qc/0407064].
\bibitem{Carter:1968rr}
  B.~Carter,
  Phys.\ Rev.\  {\bf 174} (1968) 1559.
\bibitem{Kanti:1997br}
  P.~Kanti, N.~E.~Mavromatos, J.~Rizos, K.~Tamvakis and E.~Winstanley,
  Phys.\ Rev.\ D {\bf 57} (1998) 6255
  [hep-th/9703192].
\bibitem{Cunha:2015yba}
  P.~V.~P.~Cunha, C.~A.~R.~Herdeiro, E.~Radu and H.~F.~Runarsson,
  arXiv:1509.00021 [gr-qc].
	
\bibitem{Astefanesei:2006dd}
  D.~Astefanesei, K.~Goldstein, R.~P.~Jena, A.~Sen and S.~P.~Trivedi,
  JHEP {\bf 0610} (2006) 058
  [arXiv:hep-th/0606244].
	
\bibitem{Horowitz}  
  J. M. Bardeen and G. T. Horowitz, Phys. Rev. D 60, 104030 (1999)
[arXiv:hep-th/9905099].	

\bibitem{Blazquez-Salcedo:2013muz}
  J.~L.~Blazquez-Salcedo, J.~Kunz, F.~Navarro-Lerida and E.~Radu,
  Phys.\ Rev.\ Lett.\  {\bf 112} (2014) 011101
  [arXiv:1308.0548 [gr-qc]];
		\\
  J.~L.~Blazquez-Salcedo, J.~Kunz, F.~Navarro-Lerida and E.~Radu,
  Phys.\ Rev.\ D {\bf 92} (2015) 4,  044025
  [arXiv:1506.07802 [gr-qc]].

	


\end{thebibliography}
\end{document}